\documentclass[useAMS,usenatbib]{mn2e}
\usepackage{graphicx,amssymb,amsmath,layout,verbatim,rotating,calc,mathrs
fs,natbib}
\usepackage{graphics}
\usepackage{rotate}
\usepackage{float} 
\usepackage{multirow}
\usepackage{xcolor} 
\usepackage{longtable}
\usepackage{arydshln}
\usepackage{txfonts}
\newcommand{\ej}{E$_{J}$ }

\title[non-x1-tree orbits]{The orbital content of bars:\\
The origin of ``non-x1-tree'', bar-supporting orbits}
\author[P.A. Patsis \& E. Athanassoula]
{P.A. Patsis,$^{1,2}$\thanks{patsis@academyofathens.gr (PAP)}
E. Athanassoula,$^{2}$\\
$^1$Research Center for Astronomy, Academy of Athens, Soranou Efessiou
4, GR-115 27, Athens, Greece\\
$^2$Aix Marseille Universit\'{e}, CNRS, CNES, LAM, Marseille 13, France\\
}
\date{Accepted ..........Received .............;in original form ..........}

\begin{document}
\maketitle

\label{firstpage}
 
\begin{abstract}
Recently, many orbital studies in barred galaxy potentials have revealed
the existence 
of orbits which are {\it not} trapped around x1-tree orbits, but could be 
potentially appropriate building blocks for bars. 
These findings  question the uniqueness of the x1 family as the 
standard
paradigm of orbital motion in galactic bars. 
The main goal of this paper is to investigate the role that such 
orbits could play in shaping the morphology of bars. We trace 
the morphological patterns appearing in the face-on and edge-on views of the 
non-periodic orbits presented in these studies and we show that they are 
introduced in the system by second type (\textit{``deuxi\`{e}me genre''}) 
bifurcations of x1. For this purpose, we use a typical 3D Ferrers bar
model and 
follow the radial and vertical bifurcations of the x1 family considered as being 
$mul$-periodic, with $mul=2,3,5$. The variation of the stability indices of x1 
in 
the $mul=2,3$ cases give us also the $4-$ and $6-$periodic orbits, 
respectively. 
We tabulate these orbits  including all information necessary to assess their 
role as appropriate building blocks. We discuss their stability and their 
extent, as well as their size and morphological evolution, as a function
of energy.
We conclude that even the most important of the $mul$-periodic orbits 
presented
in Tables 2 to 5 are less appropriate building blocks for  bars
than the families of the x1-tree  at the same energy.
\end{abstract}

   
\section{Introduction}
\label{sec:intro}
In general, studies of orbital structure have shown that the basic building 
blocks of 2D bars are elliptical-like,
elongated along the bar, single-periodic orbits confined within the corotation 
resonance, which are called x1 or B \citep[][etc.]{cp80,
a83, cg89}. Bars, however, are not 2D objects. They have a considerable extent
vertically to the equatorial plane of the galaxy and a rather complex shape
\citep[for a review see sect. 3.3 of][and references therein]{a16}. Building
blocks for such a 3D bar could be the families of the
x1-tree\footnote{The set
of families belonging to the x1-tree, includes x1 and its simple-periodic 3D
bifurcations at the vertical n:1 resonances.} \citep{spa02a,psa02}, with
elliptical-like projections on the equatorial plane. Many recent papers,
however, discuss new building blocks with quite different morphology. Typical
examples of such orbits can be found in \citet[][hereafter AVSD]{avsd17},
\citet{cppsm17}, \citet{dvm11}, \citet[][hereafter GLA]{ggla16},
\citet[][hereafter KPP]{kpp11}, \citet{mm16}, \citet[][hereafter PKa and PKb,
respectively]{pk14a, pk14b}, \citet[][hereafter PWG]{pwg15}, \citet[][hereafter
VSAD]{vsad16}, \citet[][hereafter WAM]{wam16} and \citet[][hereafter
WM-D]{wd09}. In some cases the face-on view of these orbits is complicated,
while their edge-on views are similar to the projections of the frown-smiles
orbits that are considered to build the peanut \citep[see e.g.][]{spa02a}. The
orbits depicted in the figures of the above mentioned studies are not periodic,
but their approximate morphology indicates that they are pretty near a periodic
one, or that they are to a large extent determined by the presence of a radial
or vertical resonance. The existence of these orbits raises the question, of
whether the paths followed by stars in galactic bars are finally different from
the generally accepted elliptical-like orbits. Elliptical-like
projections of periodic orbits on the equatorial plane of barred galaxy models, 
with their major axes aligned with the major axis of the bar, are secured by 
the 
presence of x1 and its
simple-periodic 3D bifurcations, i.e. by the orbits of the x1-tree, at least
away from the regions of radial n:1 resonances with n$>2$. 
Higher multiplicity periodic orbits have in general more complicated shapes, 
but could, in
principle,
still offer an alternative, more complicated, backbone of bars. In the
present 
paper we address this possibility in view of the new results found
recently in the above cited papers.

The periodic orbits (hereafter p.o.) found in 3D rotating bars belong to
families, which in principle exist in every 3D rotating potential in autonomous
Hamiltonian systems. Such systems can be written, in Cartesian coordinates 
$(x,y,z)$, in
the form:
\begin{equation}
H= \frac{1}{2}(p_{x}^{2} + p_{y}^{2} + p_{z}^{2}) +
    \Phi(x,y,z) - \Omega_{p}(x p_{y} - y p_{x})    ,
\label{eq:ham}
\end{equation}
where $p_{x},~ p_{y},$ and $p_{z}$ are the canonically conjugate momenta and
$\Omega_{p}$ the angular velocity of the system (pattern speed). The numerical
value of the Hamiltonian, E$_J$, is the Jacobi constant and we will also refer
to it throughout the paper as the ``energy''. We want to underline that it is 
not
the presence of a specific barred component that gives rise to these
families. Examples of models without an explicit bar component, where the same
families appear can be found in \citet{pz90}, \citet{pg96}, \citet{pags02} and
\citet{cppmp19}.
The morphology of the p.o. is determined merely by the presence of the radial
and vertical resonances \citep{spa02a}. These resonances exist in any model of a
rotating three-dimensional potential. 

Periodic orbits determine the dynamics of a dynamical system, since they determine
the topology of the phase space. Around any stable p.o. there are islands of
stability, while the unstable p.o. introduce chaotic motion. The multiplicity
$mul$ of
a periodic orbit plays a key role as it gives the number of intersections of the p.o.
with the surface of section in one direction (usually the upwards
intersections are considered). Consequently, a p.o. of multiplicity $mul$, 
called
a
$mul$-periodic orbit, will have $mul$ intersections with the surface of
section. We note that $mul$ depends on the chosen surface of section. A typical 
example is the two-dimensional (2D) case of a x1 orbit with loops at its 
apocenters
\citep[see e.g. the orbit in figure 2e in][]{spa02a}. Placing the bar major axis
along the y-axis, and considering the $y=0$ surface as surface of section, it is
1-periodic, while if we consider the surface of section $x=0$, it is 3-periodic.
Hereafter, in our orbital studies we use $y=0$ as cross section.
	
Among all p.o., the simple p.o. of multiplicity 1, or 1-periodic, play a key
role. If stable, they are found at the centers of large stability islands in the
surfaces of section. In potentials, which have a sufficiently strong bar, these
islands are surrounded by a system of smaller islands forming what is
often referred to as an archipelago. We have chains of smaller islands in
resonance zones around the main island, in which stable-unstable pairs of p.o.
alternate \citep[see e.g][]{gcobook}. These are p.o. of multiplicity $mul$ with
$mul>1$, frequently encountered in the literature in the terminology of
Poincar\'{e}, namely as type II (second type, ``\textit{deuxi\`{e}me genre}'') 
orbits
\citep{poin99}. An attempt to visualize the succession of invariant tori and the
chaotic filaments that connect the unstable p.o. in the 4D space of section of
3D Hamiltonian systems has been presented in KPP. 

Tracing the \textit{deuxi\`{e}me genre} orbits is not always an easy task,
especially if the dimension of the system is larger than two. The standard tool
frequently used for finding the families of periodic orbits and their connection
with other bifurcating families in 3D Hamiltonian models, is the ``stability
diagram'', which gives the linear stability of the members of a family of p.o.
as a parameter, (e.g. \ej\!\!), varies \citep{cm85}. For calculating the linear
stability of the periodic orbits in such systems we proceeded as \citet{b69} and
\citet{hdj75}. Definitions and details about the algorithm we followed can be
found e.g. in \cite{spa02a}. Here we briefly mention that the variation of the
stability with \ej is described by the variation of two stability indices, $b_1$
and $b_2$, one of which refers to the stability of the orbits of a family
subject to radial and the other to vertical perturbations. A p.o. is stable (S)
if both $b_i\in(-2,2)$, with $i=1,2$. If one of the two stability indices is
$|b_i|>2$, then the orbit is characterized as simple unstable (U), while if both
indices are $|b_i|>2$ it is double unstable (DU). Finally if all four
eigenvalues of the monodromy matrix are complex numbers \textit{off} the unit
circle, the stability indices cannot be defined and the p.o. is called complex
unstable $(\Delta)$. In the case of 1-periodic orbits, whenever we have
intersections or tangencies of one of the $b_i$ stability curves with the $b=-2$
axis, new families are bifurcated from the parent one, having the same
multiplicity. On the other hand, tangencies or intersections with the $b=2$ axis
bring into the system new families with multiplicity double that of 
the parent one \citep{gco86}. 

In general, the critical value of a stability index at which a bifurcating
family of multiplicity $mul$ is introduced in the system is given by 
\begin{equation}
b = - 2\cos\left( 2 \pi \frac{1}{mul}\right). 
\label{eq:b}
\end{equation}
Thus, a family of p.o. of multiplicity $mul=1$, reaching the $b=-2$ axis as \ej
varies, will bifurcate another family again with $mul=1$, while families with 
$mul=2$
will appear, according to Eq.~\ref{eq:b}, at intersections or tangencies of a
stability curve with the $b=2$ axis. A stability diagram of a 1-periodic family
gives us the information about the \ej value at which 1- and 2-periodic families
will be introduced in the system. However, for tracing further families with
$mul=3,4,5$ etc. in the same diagram, we have to plot, besides the  $b=-2$ and
$b=2$ axes, also the axes $b=1,0,-0.618...$ and so on respectively \cite[see
Appendix A in][for details]{spa02b}.

An alternative way for following $mul$-periodic bifurcations of an 1-periodic
family is to consider it as being itself of multiplicity $mul$, by repeating it
$mul$ times. Then, the stability curves will indicate the birth energies of the
bifurcating families of multiplicity $mul$ at their intersections, or
tangencies, with the $b=-2$ axis, while new families with multiplicity $2mul$ 
will
be introduced in the system, whenever $b=2$. Hereafter, if not otherwise
indicated, when we refer to $mul$-periodic orbits, we mean that $mul \geq 2$.

The goal of this paper is to find the origin of the orbital 
morphologies
encountered in relevant recent papers on orbital structure cited in the first
part of this introduction. In this effort we have to have in mind two things:
Firstly that the shape of an orbit is influenced by the resonance in which it is
introduced in the system, but also by the energy along the 
characteristic\footnote{The characteristic of a family of p.o. is a curve that 
gives one (or more) initial condition of its members as a 
function of a parameter, usually \ej.} at
which a particular representative of a family is found. As an example one can
observe the morphological evolution of x1 along its characteristic in various
models, like figure~11 in \citet{gco83}, figure~3 in \citet{a92}, or figure 2 in
\citet{spa02a}. Secondly, that moving along the characteristic of
a 3D family, each projection of a p.o. will follow its own morphological
evolution.
On top of this, the face-on and edge-on projections of the (not necessarily
periodic) orbits may display the fingerprints of different radial and vertical
resonances in their morphologies, depending on the location of its initial
conditions in phase space. A certain face-on structure may be combined with
different edge-on ones and vice versa. Thus, when we look for a specific face-on
or edge-on morphology, essentially we want to identify the energy at which this
specific structure is introduced in the system.  

In our study we consider also the possibility that chaotic orbits may support
the bar, or part of it. The role of chaotic orbits in supporting a bar has been
investigated in
several studies of 2D and 3D models \citep[see e.g.][WAM]{a83, pkg10, ma11,
mm16}. Note that PKa found that chaotic orbits close to the transition
points from stability to complex instability \citep{cm85} may
contribute in the reinforcement of features like the X's in edge-on views of the
bars. In addition sticky chaotic orbits \citep[][see also figure 8 and 
corresponding text in Athanassoula et al. 1983]{ch08} at the vertical 2:1
resonance (vILR) region (PKa, PKb), as well as sticky orbits in the
outer regions of specific 3D bars \citep{cppsm17} seems to favor the appearance
of boxy features in the inner and outer regions of the bars respectively. Sticky
chaotic orbits have been also found to shape the bars in studies of 2D models
\citep{paq97, a10}.


The question that we try to answer in this paper is whether or not there are
families of periodic orbits (other than the x1-tree set of families) that
are associated with the reinforcement of the bar and, in particular, of the
morphological features we observe in their central parts (i.e. peanuts and X
features embedded in them). For this purpose we study the \textit{``deuxi\`{e}me
genre''} periodic orbits in Ferrers bars\footnote{These ellipsoids have been
already presented in the 19$^{th}$ century by \citet{f887} and have been used
extensively in Galactic Dynamics, first introduced by \citet{a83}.} and we
compare their morphologies to those appearing in the papers mentioned at the
beginning of the introduction. We do not intend to present in detail the orbital
properties of the families. We rather want to point to their origins and discuss
whether they could in principle be  part of the skeletons of the bars.
In order to trace the \ej at which a $mul$-periodic family is ``born'',
we follow the evolution of the stability indices of x1, as we did in previous
works \citep[][PKa,b etc.]{spa02a}. For the purposes of the present paper,
however, we follow the rules described earlier, and consider x1  each time as
$mul$-periodic.  We list all the families we found in this way in tables, and we
note that their vast  majority is presented for the first time.  The few cases
that correspond to families of p.o. known from earlier works, are accordingly
indicated in these tables.

This is the first in a series of papers about the orbital content of galactic
bars, in which our main goal is to point out the origin of the various patterns
and locate the energies at which they are introduced relatively to the main
radial and vertical $n:1$ resonances and especially relatively to the vILR. In
Section~\ref{modelorbs} we describe the model, the particular set of 
parameters
we have chosen and the nomenclature we use for the orbits. In 
Section~\ref{sec:2po} we present the main 2- and 4-periodic
orbits, in Section~\ref{sec:3po} the main 3- and 6-periodic ones, while in
Section~\ref{sec:highej} we investigate the role of $mul$-periodic orbits with
large energies in determining the structure of the bar. Finally in
Section~\ref{sec:concl} we present and discuss our conclusions.

In Appendix~\ref{sec:qporbs} we present morphologies appearing in quasi- and
non-periodic orbits and we discuss extensively their similarity with orbits
recently found in many models in the relevant literature. These models are
either analytic, or have been obtained from snapshots of N-body simulations. In
the Appendix we do not present any further new family of periodic orbits, but we
compare orbital patterns and we suggest connections between the shapes of
published non-periodic orbits and the $mul$-periodic orbits we present in the
tables in the main body of the paper. However, readers who just want to read the
bottom line of our work should concentrate on Sections~\ref{sec:2po},
\ref{sec:3po}, ~\ref{sec:highej} and \ref{sec:concl}, or even go directly to
Section~\ref{sec:concl}.

\section{Introducing the models and the orbits}
\label{modelorbs}
\subsection{models}
\label{models}
For the sake of continuity with our previous studies on the subject, we use
again
in this paper the popular triaxial Ferrers bar model, which is described in
detail in 
\citet{spa02a} and \citet{pk14a}, with parameters close to those in the pioneer
paper by \citet{pf84}. The formulae for the axisymmetric part of the potential,
as well as the bar model can be found in these references.

The Ferrers
bar is inhomogeneous with index 2. He have taken as its major axis our $y-$axis
and axial ratios $a:b:c = 6:1.5:0.6$ ($a$, $b$, $c$ are the semi-axes with
$a>b>c$). 

The axisymmetric background consists of a Miyamoto disc \citep{mn75} with fixed
horizontal and vertical scale lengths A=3 and B=1 respectively and a Plummer
sphere \citep{pl11} for the bulge with scale length $\epsilon_s$.

The length unit is taken as 1~kpc, the time unit as 1~Myr and the mass unit as $
2\times 10^{11} M_{\odot}$. The masses of the three components satisfy \(
G(M_{D}+M_{S}+M_{B})=1 \), where \( M_{D} \) is the total mass of the disc,
\(M_{S} \) the mass of the bulge (spheroid), \( M_{B} \) is the mass of the
bar component and G the gravitational constant.

In all papers mentioned in the beginning of this section one can find detailed 
descriptions of the components of the models and their properties, so we skip 
this information here.

\subsection{Radial and vertical perturbations and the choice of convenient
parameters}
\label{params}
In this paper we use two specific models, both following the description in the 
previous subsection, but with different values for the parameters. The first 
one, used only briefly in this subsection, is the fiducial model of 
\citet{spa02a}, i.e. has  
\( GM_{D}=0.82,
GM_{S}=0.08, GM_{B}=0.1, \epsilon_s =0.4 \). The frame of reference rotates
with the bar pattern speed $\Omega_{p} = 0.054$, which places corotation at a 
radius $R_c=6.13$ from the center.

\begin{figure}
\begin{center}
\resizebox{80mm}{!}{\includegraphics[angle=0]{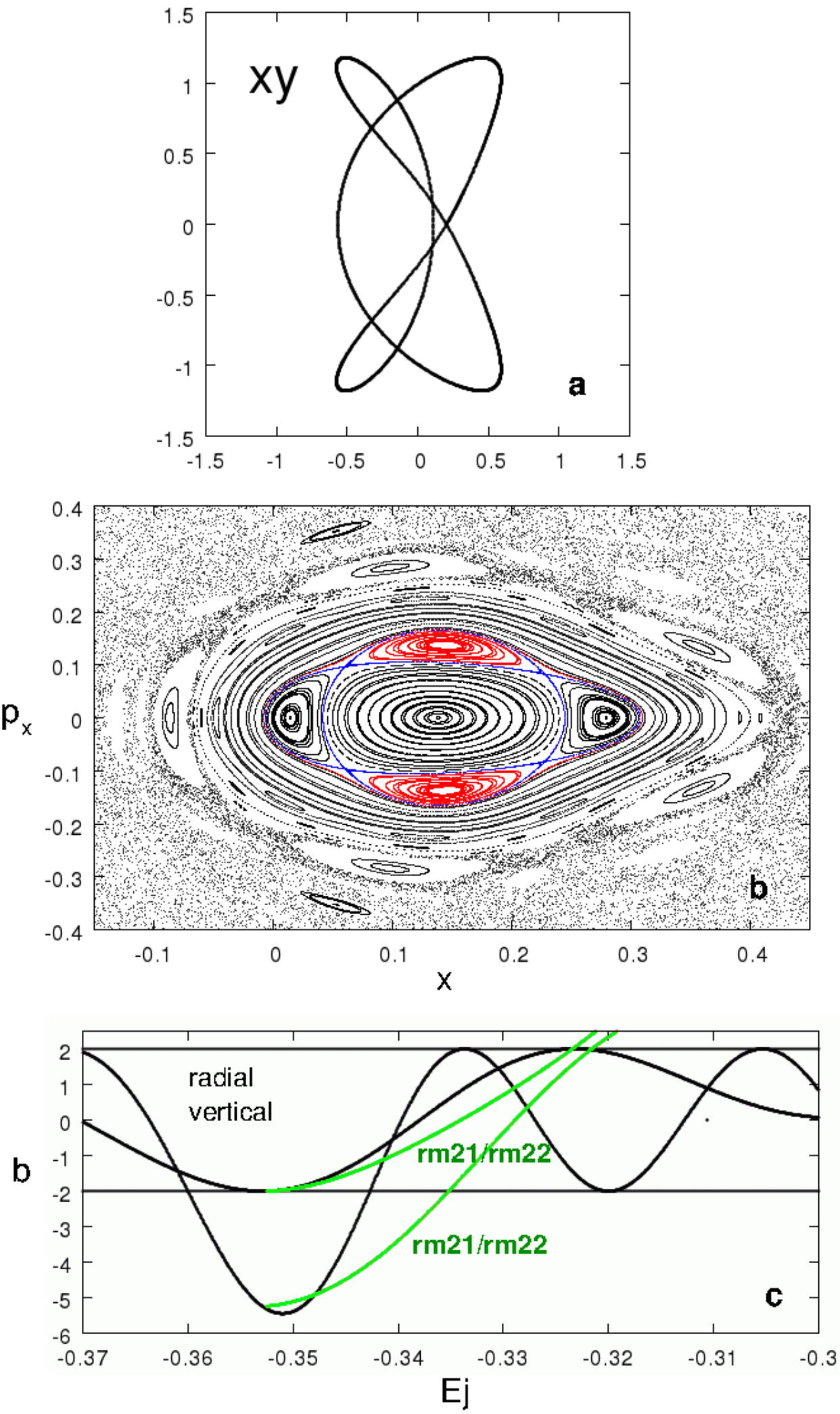}}
\end{center}
\caption{(a) A portrait of an rm22 orbit in the fiducial model of
\citet{spa02a}. (b) The $(x,p_x)$ Poincar\'{e} surface of section close to x1 at
\ej$=-0.348$
in the same model. (c) The stability curves of x1 (black lines) and rm21 and 
rm22 (green
lines). We observe that for \ej$=-0.348$ the rm21/rm22 families, sharing the 
same
stability curves, are vertically unstable but radially stable. Thus in (b) they
are found in the middle of stability islands coming in pairs, above and below
(red) and left and right (black) of the central island that belongs to x1.}
\label{ofmainm2} 
\end{figure}

By searching for the morphologies of the orbits in the papers mentioned in the
introduction, we realized that an orbit frequently encountered in the fiducial
and in
many other Ferrers bar models we have investigated \citep[see e.g.][]{pk14a} was
the
planar orbit depicted in Fig.~\ref{ofmainm2}a. In this and in all subsequent
plots
with orbits the numbers in the axes refer to kpc. In the plots of orbits,
usually in
the upper left corner, we indicate the depicted projection.
The orbit in  Fig.~\ref{ofmainm2}a belongs to a known family of p.o. that is
introduced in the system at the first tangency of the \textit{radial} stability
index
of the simple-periodic x1 family with the $b=2$ axis and thus it is 2-periodic.
This
family has been discussed in PKb (figure 2 in that paper) and has two branches
named rm21 and rm22,
because it comes in pairs symmetric with respect to the major axis of the bar.
On the
$(x,p_x)$ Poincar\'{e} surfaces of sections one can always observe two sets of
stability islands belonging to these orbits. For example Fig.~\ref{ofmainm2}b
depicts
the surface of section of the fiducial model for \ej$=-0.348$, i.e. an energy
right
after the bifurcation of rm21 and rm22 from x1. The x1 p.o. is located in the
middle of the
central stability island, at $(x,p_x)\approx (0.138,0)$, while the two sets of
islands in its immediate neighborhood belong to rm21 (left and right of x1) and
rm22
(above and below x1, plotted in red). If we plot together both members of this
family in the same figure, we observe a rather boxy shape \citep[see figure 2
in][and
also Section~\ref{sec:2po} below]{pk14b}.

The families rm21 and rm22 are important and play a major role in
explaining boxiness
in the central region of the bar \citep{pk14b, cppsm17}. Similar morphologies
are
also encountered in the face-on views of the orbits in figure 7 (third and
fourth
row) in GLA, in figure 2 in KPP, in figure 4 (1st row) in VSAD and in figure 5
(third
column) in WAM. We come up against this morphology almost in all studies about
orbits
in rotating galactic bars.

In Fig.~\ref{ofmainm2}c we give the stability diagram of the x1 family 
described twice, i.e.
when we consider it as being 2-periodic. The stability indices of x1 are plotted
with
black lines. Arrows indicate which curve corresponds to the radial and which one
to
the vertical index. Bifurcations of equal multiplicity will occur at tangencies
or
intersections of the stability curves with the $b=-2$ axis and will be
2-periodic in
this case. We observe that rm21 and rm22 are bifurcated at \ej $\approx -0.353$
(green
stability curves) at the tangency of the x1 radial stability curve with the
$b=-2$
axis. Both share the same stability curves, thus both are bifurcated as simple
unstable and then they have a U $\rightarrow$ S
$\rightarrow$ U $\rightarrow$ DU transition, in which the stable part
$\Delta$\ej$\approx 0.012$. The fact that the pair of rm21 and rm22 is simple
unstable does not
contradict the presence of invariant curves around it in Fig.~\ref{ofmainm2}b.
Its
radial index is $-2<b<2$ indeed, while the vertical one is $b<-2$, starting from
the
corresponding index of x1. The \ej at which rm21, rm22 are bifurcated is in the
vILR
region, where we have for x1 the S $\rightarrow$ U $\rightarrow$ S transition,
roughly for energies $-0.36$ and $-0.342$, at which first the x1v1 and then the
x1v2
families are introduced in the system \citep{spa02a}. This means that vertical
perturbations of rm1,2 will correspond mainly to orbits belonging to a chaotic
sea,
so it would be difficult to compare the side-on morphologies with those
presented in
the relevant literature. Such issues can be raised in several models and that
led us
to seek a Ferrers bar model in which the basic families under discussion existed
and
were introduced as stable. Thus, we performed a preliminary investigation of the
energies at which the main 2-, 3- and 4-periodic families are introduced. We
have
chosen to present the basic $mul$-periodic families encountered in rotating 3D
bars in
a model close to the fiducial one, which however has $GM_{S}=0.022$ instead of
$GM_{S}=0.08$. Thus, as both the mass of the bar and the total mass of the
system
remain constant, the disc mass increases. The parameters of the main model of
this
paper are summarized in Tab.~\ref{tab:param}.
\begin{table}
\caption[]{The parameters of the main model. G is the
  gravitational constant, M$_D$, M$_B$, M$_S$ are the masses of the
  disk, the bar and the bulge respectively, $\epsilon_s$ is the scale
  length of the bulge, $\Omega_{b}$ is the pattern speed of the bar.}
\label{tab:param}
\begin{center}
\begin{tabular}{ccccccccc}
  GM$_D$ & GM$_B$ & GM$_S$ & $\epsilon_s$ & $\Omega_{b}$  \\ 
\hline
  0.878 &  0.1  & 0.022 & 0.4 &  0.054  \\

\hline
\end{tabular}
\end{center}
\end{table}
In models with bars heavier than that of the fiducial \citep[see figure 3
in][]{pk14b}, as well as in models with heavier classical bulges, the rm21,
rm22 families
start existing between the bifurcating points of x1v1 and x1v2, as vertically
unstable.

Useful quantities for assessing the role of the orbits we discuss in the
following
sections are the (\ej\!\!,$y$)=$(-0.197,6.2)$ coordinates on the characteristic
diagram of the unstable Lagrangian point $L_1$ for $x=0$ and the
\mbox{(\ej\!\!,$x$)}=$(-0.193,5.9)$ coordinates of the stable Lagrangian point
$L_4$
for $y=0$. We also note that the longest x1 orbit reaches $y=4.1$~kpc along the
major
axis of the bar.

\subsection{Orbital nomenclature}
\label{names}
Families commonly used in previous studies -- such as x1, x1v1, etc. --
already have a name so we keep it also in the present study. In the tables in
which we classify the p.o. (sections~\ref{sec:2po} and \ref{sec:3po}) we give,
for reasons of continuity and clarity, in parenthesis also the name that these
families would have if we followed the rules we describe below. In the text,
however, we give only the standard name, for brevity.

For the remaining families, we introduce the following nomenclature: The first 
character indicates whether the family is introduced in a radial (r) or a 
vertical (v) bifurcation of the x1. This is followed by two characters 
indicating its multiplicity, e.g. m2 or m3, and a third character which is a 
number giving the order in which this family is introduced in the system. For 
the unstable counterparts of the stable $mul$-periodic families that are 
bifurcated at the same energy, we add a ``u'' at the end of the name, retaining 
the rest of it identical.  So e.g. the name rm21 indicates that this is the 
first radial bifurcation of x1 of multiplicity 2. This family is bifurcated 
together with the rm22 family, with which it shares the same stability curves 
and the orbits of which are symmetric of the rm21 orbits with respect to the 
y-axis. At the same energy we also have the bifurcation introducing the pair of 
rm21u and rm22u families, which are the unstable counterparts of rm21, rm22. 
Since we need a name mainly to associate it with a certain morphology, the 
clarification is finally done in our Tables, where names and morphologies are 
presented side-by-side.

Note that a family of multiplicity $mul$ will appear also as $2mul$, $3mul$ etc.
Thus,
we keep for the latter the name of the lowest multiplicity with which the family
is encountered in the stability diagrams, in order to avoid multiple names for
the same family.

Finally, we note that we found only very few important vertical or radial 
bifurcations from the x1-tree families. So there was no need to extend the 
nomenclature rules to include such orbits. In the few cases we mention in the 
sections below we used names that clearly show their origin from the parent 
family. As an example the family r$\_$tv1 indicates that it is the first 
vertical bifurcation in the set of radial 3:1 bifurcating from x1 families, for 
which we used in previous papers names starting with ``t'' \citep{spa02a}.

In many studies a family is named by the ratio of its frequencies, either in
Cartesian or cylindrical coordinates. Although this is often quite adequate for
many dynamical studies, it is definitely inadequate for orbital structure work,
as several families with quite different properties (such as energy ranges,
stability, extents and shapes) have the same frequency ratios. Even
naming the orbital families just by the name of the
resonance at which they appear, as is usually done for rotating disks,
might
lead to ambiguities.  A good example is that of the x1 and x2 families, both of 
which
would, by the above definition, be named 2:1.  We have thus refrained from using
names based on
frequencies and/or resonances here.

\section{2-periodic orbits}
\label{sec:2po}
\begin{figure*}
\begin{center}
\resizebox{160mm}{!}{\includegraphics[angle=0]{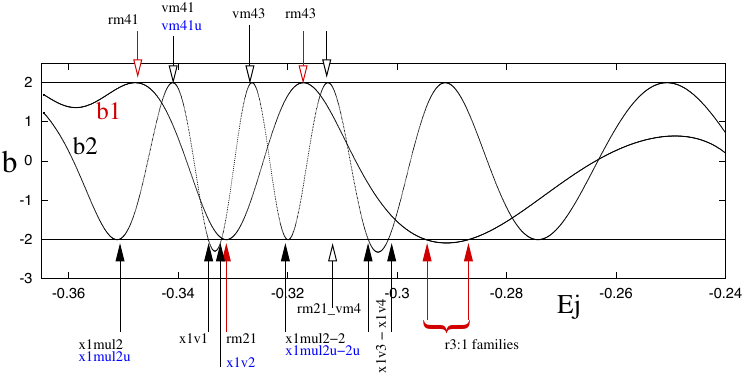}}
\end{center}
\caption{The stability diagram of the x1 family considered as 2-periodic. The
index b1 refers to radial, while b2 to vertical perturbations. Red arrows
point to the energies at which 2D families are introduced in the system, while
black to the corresponding energies for 3D families. Note that the
families
indicated at the tangencies of the stability curves with the $b=2$ axis, are
introduced as 4-periodic. The names of the families
in blue indicate that they are introduced as unstable.}
\label{x1stm2} 
\end{figure*}

Having chosen a convenient model -- i.e. a model in which the
rm21/rm22 families exist, are bifurcated as stable and span a
sufficient extent -- we proceed with finding the families of periodic
orbits.
In our study we restrict ourselves to the bifurcations of x1 and 
we do not
include bifurcations of the z-axis family 
\citep{hms}, as these are not relevant to our study. In order to present the 
orbits in
an illustrative way, we consider the
family x1 as 2-periodic and we obtain the stability diagram given in
Fig.~\ref{x1stm2}, which shows the origins of all families we discuss in
Section~\ref{sec:2po}.
We avoid giving the evolution of the stability indices of all these families in 
Fig.~\ref{x1stm2} or in separate figures, because we focus mainly in the 
succession in which these families are introduced in the system. The extent of 
the families and their possible interesting bifurcations, are discussed   
whenever they play a role. Moreover we restrict ourselves to a large extent in 
presenting the families, whose size and shape  could make them potential 
contributors to the bar.

The main 2-periodic families that could play a role in building side-on peanut
structures are summarized in Table~\ref{tab:mul2tab}. In our presentation we give
only one representative of each family. However, given the 4-fold symmetry of our
adopted potential, there exist also symmetric orbits, like e.g. the ``frown'' and
``smile'' orbits of the x1v1 and x1v1$^\prime$ pair, which are symmetric with respect
to the equatorial plane. The role of symmetric orbits in building observed structures
will be discussed in Sect.~\ref{sec:concl}. Whenever we give initial conditions we
give the $(x,z,p_x,p_z)$ array, since we use the $y=0$ surface of section. Numbers in
initial conditions are given in the text at least with a three decimal digits
accuracy, so that the orbit can be easily recovered with an iterative scheme. In the
present study the p.o. are considered to be found when the initial and final
coordinates coincide with an accuracy of at least $10^{-11}$, using a fourth order
Runge-Kutta method. The relative error in the energy is always less than $10^{-15}$.

Table~\ref{tab:mul2tab} includes all data we need in order to assess the role of each
family in a compact manner. In the first column we mention the name with which we
refer to the family in our analysis, while in the second column the main information
given is the energy E${_J}{^*}$, at which the family is born (cf. with
Fig.~\ref{x1stm2}). We also add some description about the evolution of the stability
of the family. The most important columns are the three that follow, where one can
directly associate successively the face-on, end-on and side-on projections of the
orbits with the name given in the first column. Stable orbits are plotted in black,
while unstable ones in blue. Finally in the last column, we give information about
the specific representative of the family depicted in the third, fourth and fifth
columns. We give its energy and approximate initial conditions. We always chose a
representative of the family that is characteristic over a large $\Delta$\ej
interval. For example if a family supports a specific face-on profile for a large
energy interval, we will present it at an energy within this interval, away from the
bifurcating point, where it will be x1-like anyway. Also in the last column we give
reference to other works, where similar morphological patterns appear. For the main
papers we compare our study with, we use the abbreviations defined in the
introduction. Indicating that a p.o. orbit in Table~\ref{tab:mul2tab} is ``similar''
to an orbit found in the literature, we mean that itself, or a quasi-periodic orbit
in the immediate neighborhood within its stability island, or a chaotic orbit 
sticky to
it \citep{ch08}, can be considered as morphologically similar to the orbit in the
cited paper.  We included only examples where we have conspicuous similarities.
However, there are more cases that could be included if we take into account changes
that appear as we move along the characteristic of a family \citep[cf. for example
the orbit rm21u with the face-on view of the orbit in figure 11
in][]{dvm11}.

The orbits are presented in order of increasing E${_J}{^*}$ from the top to the
bottom of the tables (cf. with Fig.~\ref{x1stm2}). The bifurcations of the
bifurcating families do not play in general a major role. In the tables we
include only those that we find that could be important for supporting the 3D
bar structure. In such a case they are given below the parent family separated
by a dashed line (see the case of family rm21\_vm4 in Table~\ref{tab:mul2tab}). 

Special mention has to be given to the group of 3:1 orbits that one can start tracing
at the S$\rightarrow$U transition of x1 at E${_J}{^*}\approx -0.295$ (indicated
with
``r3:1'' in Fig.~\ref{x1stm2}). It includes families with stable and unstable parts
and face-on morphologies similar to the p.o. t1, t2 and o1 in \citet{spa02a}. The
tree of 2D and 3D families starts with a \textit{radial} inverse bifurcation
\citep{cm85} and extends towards larger as well as towards smaller
energies than its E${_J}{^*}$, with
branches joining again the x1 as well as the x4 characteristics. The interconnections
of all their stability curves are complicated and a detailed description is beyond
the scope of the present paper. We restrict ourselves to the presentation of a
representative of a 3D family with stable parts, which is given in
Table~\ref{tab:mul2tab} with the name r\_tv1, mainly because of its side-on profile.
This profile could be possibly used for building boxy or X-shaped side-on
projections. 

In Table~\ref{tab:mul2tab} we have also included the simple periodic (of
multiplicity 1) x1v1 and x1v2 families, as essentially the contribution of all
the remaining families is discussed in comparison with the edge-on profiles that
can be built by mechanisms using these two families as building blocks. It has
to be added that in the particular model we use for the presentation of the
orbits, the x1v1 family does not have complex unstable parts, as in all other
known cases up to now, thus its role for building an X-feature can in principle
be based solely on quasi-periodic orbits trapped around stable p.o. This fact
upgrades the role of x1v1 and x1v2 families as peanut building blocks and
deserves further investigation, since the parameter that differentiates the
specific model we use for the presentation of orbits here with respect to other
Ferrers bar models, is the increase of the disc mass at the expense of the
bulge's mass.

\begin{figure}[h]
\begin{center}
\resizebox{80mm}{!}{\includegraphics[angle=0]{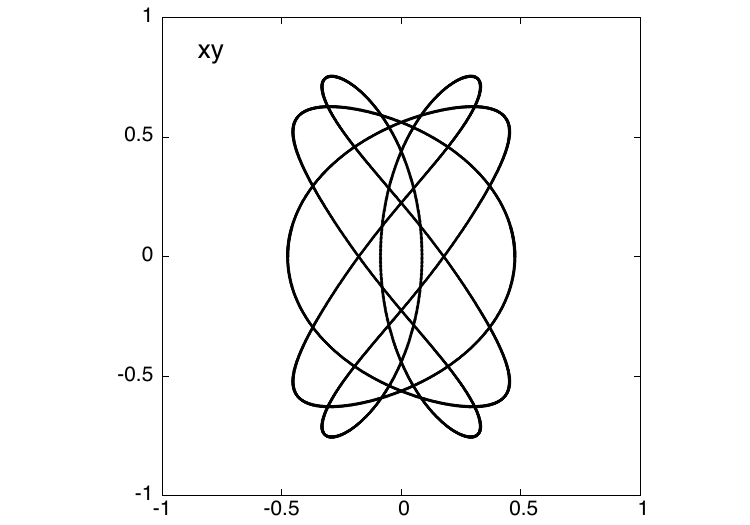}}
\end{center}
\caption{The representatives of the families rm21 and rm22 at \ej $= -0.324$ plotted
together. They can obviously be the backbone for boxy features in the regions 
they
exist.}
\label{boxy1} 
\end{figure}  
The easiness with which a pattern is encountered in different models is
reflected in
the number of works in which it is found and mentioned in the last column of  
Table~\ref{tab:mul2tab}. This can be used as a rule of thumb for the importance
of each morphological
pattern. The large frequency with which the pairs of rm21/rm22 and rm21u/rm22u
are
encountered in the studies is a result of the phase space structure of the
x1 stability islands in the inner parts of any rotating bar model. The $(x, p_x)$
Poincar\'{e} surface of section given in Fig.~\ref{ofmainm2}b is typical of the
situation. The islands of the multiplicity 2 orbits surround invariant curves
around x1, they occupy a considerable area of the large x1 stability island and they
are located in a zone with orbits that have relatively small deviations from
the initial conditions of the x1 p.o. Furthermore, their shape is indicative 
of
the morphology of the quasi-periodic orbits one can find on the neighboring invariant
curves (see also Section~\ref{sec:qporbs}). The rm21/rm22 pair offers also a
characteristic image of the skeleton of the orbits that shape inner boxiness in
models of bars. The pattern resulting when both rm21 and rm22 are considered in our
model is given in Fig.~\ref{boxy1} (cf with figure 2 in PKb). We have however,
to
bear in mind that these orbits cannot be used for shaping the overall
structure
of the bar. Bifurcated from x1, they start increasing their projections on the minor
and decreasing their projections on the major axis of the bar. As the energy, and
thus their perimeter, increases, they develop loops that extend to the sides of the
main ellipse. Namely, with increasing energy we have an evolution in the direction
from the rm21 orbit depicted in Table~\ref{tab:mul2tab} to the $(x,y)$ projection of
rm21\_vm4 given just below it in the same table. Its stable part along its
characteristic, together with that of its bifurcating families, extends up to \ej
$\approx -0.312$. At this energy the rm21 and rm22 orbits are already much
rounder than the
orbits in Fig.~\ref{boxy1}. 
Their maxima along the x- and y-axes evolve from $(x_{max},
y_{max}) = (0.18, 0.72)$, to $(0.47,075)$, to $(0.68, 0.79)$ to $(0.78,0.78)$
for \ej $= -0.331$, $-0.324$, $-0.316$ and $-0.312$, respectively.
This morphological evolution, combined with the fact that
these orbits are generated at the ILR region, relates them directly with the effect
of inner boxiness observed in many barred galaxies \citep{ab06, ed13}, in
agreement with
results of previous studies \citep[PKb,][]{cppsm17}.

\begin{table*}
\caption{The main periodic orbits bifurcated from the x1 family considered as
being 2-periodic (Fig.~\ref{x1stm2}). The successive columns give the name of
the family, following the rules described in Sect.~\ref{models} (1), the energy
at which it is introduced in the system (E${_J}{^*}$) and information about its
stability (2), the face-on, end-on and side-on views (3-5), while in the last
column (6), labeled ``comments'', we give the \ej value of the specific orbit
plotted, its approximate $(x,z,p_x,p_z)$ initial conditions and we mention
figures in papers in which similar morphologies are presented. From top to
bottom, the families are given in order of increasing energy at which they are 
introduced in the
system. Bifurcations of the families, if given, are separated from the parent
family with a dashed line. Orbits drawn with black (blue) lines are stable 
(unstable).}
\label{tab:mul2tab}
\centering
\begin{tabular}{ l  c  c  c  c  c  }
family name & E${_J}{^*}$ & face-on &  end-on & side-on & comments\\
\hline
\parbox[l]{1.8cm}{\textbf{x1mul2} \\ \textbf{(vm21)}}& 
\parbox[c]{2.0cm}{$-0.351$ \\ (1st tangency of b2 with the $b=-2$ axis). S for
\ej$< -0.283$.} & 
\parbox[c]{2.5cm}{\includegraphics[height=2cm]{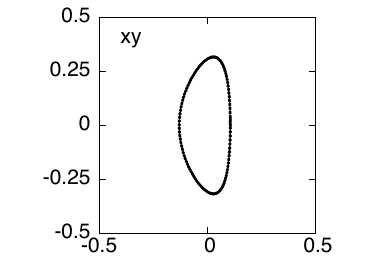}} &
\parbox[c]{2.5cm}{\includegraphics[height=2cm]{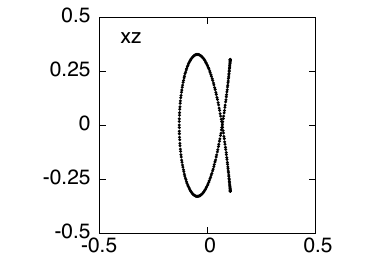}} & 
\parbox[c]{2.5cm}{\includegraphics[height=2cm]{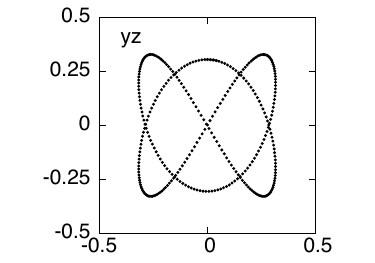}} & 
\parbox[c]{2.cm}{\ej$=-0.333$ \\ $(0.107, 0.305, 0, 0)$ \\ Figure 16
top in PKb.} \\
\hline
\parbox[l]{1.8cm}{\textbf{x1mul2u} \\ \textbf{(vm21u)}}& 
\parbox[c]{2.0cm}{$-0.351$ \\ (1st tangency of b2 with the $b=-2$ axis). U 
always.
Radially stable for \ej$< -0.292$.} & 
\parbox[c]{2.5cm}{\includegraphics[height=2cm]{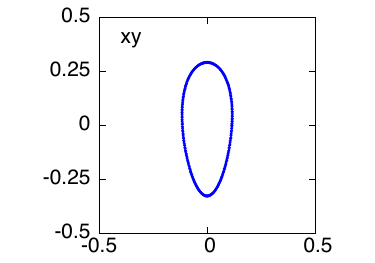}} &
\parbox[c]{2.5cm}{\includegraphics[height=2cm]{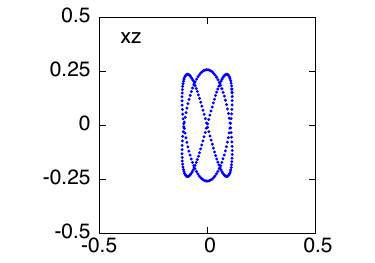}} & 
\parbox[c]{2.5cm}{\includegraphics[height=2cm]{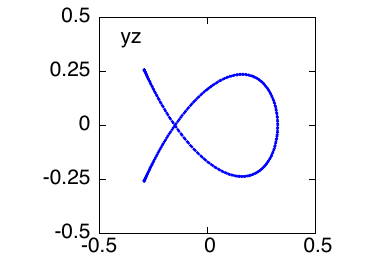}} & 
\parbox[c]{2.cm}{\ej$=-0.34$ \\ $(0.115, 0.167,$ \\ $0.008,-0.109)$ \\
\mbox{Side-on view} similar to figure 2, top, in PWG. } \\
\hline
\parbox[l]{1.8cm}{\textbf{x1v1} \\ \textbf{(vm11)}}&
\parbox[c]{2.0cm}{$-0.334$ \\ 1st vILR (frown-smile) family. S always (no $\Delta$
part in this model).} & 
\parbox[c]{2.5cm}{\includegraphics[height=2cm]{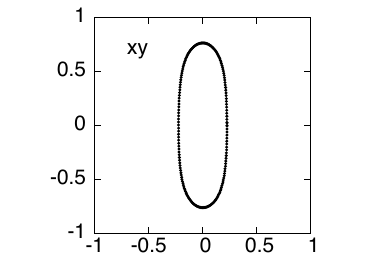}} &
\parbox[c]{2.5cm}{\includegraphics[height=2cm]{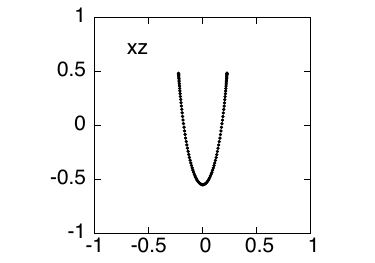}} & 
\parbox[c]{2.5cm}{\includegraphics[height=2cm]{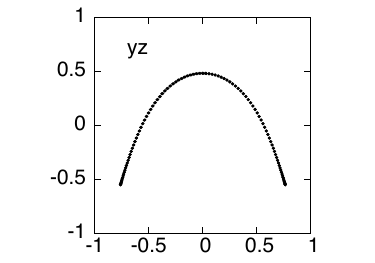}} & 
\parbox[c]{2.cm}{\ej$=-0.3$ \\ $(0.224 0.483 .0 .0)$ \\} \\
\hline
\parbox[l]{1.8cm}{\textbf{x1v2} \\ \textbf{(vm11u)}}&
\parbox[c]{2.0cm}{$-0.332$ \\ 2nd vILR ($\infty$-type) family. U always, being
radially stable.} & 
\parbox[c]{2.5cm}{\includegraphics[height=2cm]{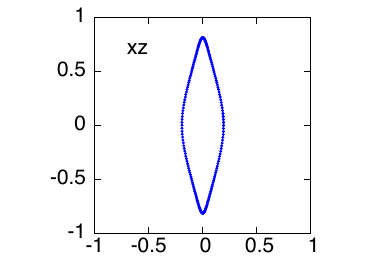}} &
\parbox[c]{2.5cm}{\includegraphics[height=2cm]{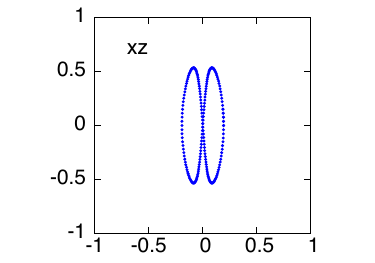}} & 
\parbox[c]{2.5cm}{\includegraphics[height=2cm]{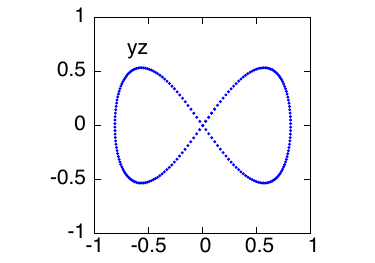}} & 
\parbox[c]{2.cm}{\ej$=-0.3$ \\ $(0.193, 0, 0, 0.275)$ \\} \\
\hline
{\textbf{rm21}}& 
\parbox[c]{2.0cm}{$-0.331$ \\ (1st tangency of b1 with the $b=-2$ axis). S for
\ej$<-0.316$.} & 
\parbox[c]{2.5cm}{\includegraphics[height=2cm]{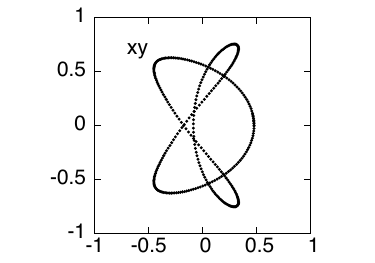}} &
  & 
  & 
\parbox[c]{2.cm}{\ej$=-0.324$ \\ $(0.475, 0, 0, 0)$ \\ Figure 2 in PKb;
similar to face-on views of orbits in figure 7 (3rd and 4th rows) in GLA,
figure 4 (1st row) in VSAD, figure 5 (third column) in WAM.\\} \\
\hdashline  
{\textbf{rm21\_vm4}}& 
\parbox[c]{2.0cm}{
\vspace{0.25cm} $-0.315$ \\4-periodic, 3D bifurcation of rm21 at
an intersection of b2 index of x1 with the $b=-2$ axis. \\S: \ej$<-0.311$.
Found as U, DU also at smaller \ej  (regression of characteristic).} & 
\parbox[c]{2.5cm}{\includegraphics[height=2cm]{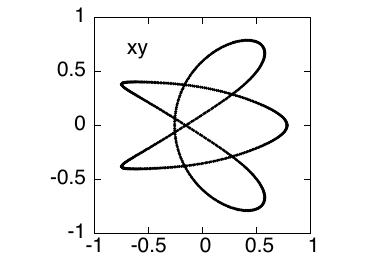}} &
\parbox[c]{2.5cm}{\includegraphics[height=2cm]{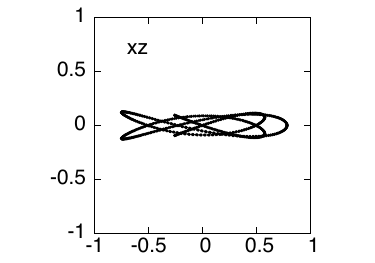}} & 
\parbox[c]{2.5cm}{\includegraphics[height=2cm]{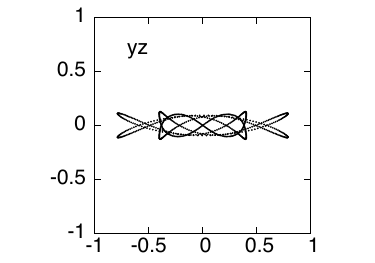}} & 
\parbox[c]{2.cm}{\ej$=-0.311$ \\ $(0.781, 0, 0, 0.048)$ \\Face-on view as for
rm21. 
}\\
\hline
{\textbf{rm21u}}& 
\parbox[c]{2.0cm}{$-0.331$ \\ (1st tangency of b1 with the $b=-2$ axis).
Initially U for \ej$<-0.316$. Then S towards smaller \ej.\\ }
  & 
\parbox[c]{2.5cm}{\includegraphics[height=2cm]{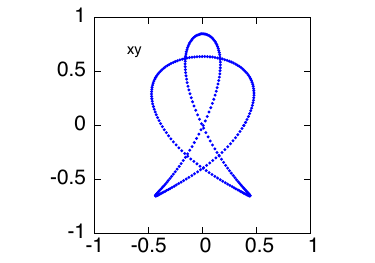}} &
  & 
  & 
\parbox[c]{2.cm}{\ej$=-0.323$ \\ $(0.381, 0, 0.102, 0)$ \\ Similar to
face-on views of orbits in figure 7 in GLA,
figure 4 in VSAD and figure 5 in WAM.\\} \\
\hline
\end{tabular}
\end{table*} 
  
\begin{table*}
\contcaption{ }
\centering
\begin{tabular}{ l  c  c  c  c  c  }
family name & E${_J}{^*}$ & face-on &  end-on & side-on & comments\\
\hline
\parbox[l]{1.8cm}{\textbf{x1mul2-2} \\ \textbf{(vm23)}}& 
\parbox[c]{2.0cm}{-0.320 \\ (2nd tangency of b2 with the $b=-2$ axis).\\ S:
\ej$<-0.236$} & 
\parbox[c]{2.5cm}{\includegraphics[height=2cm]{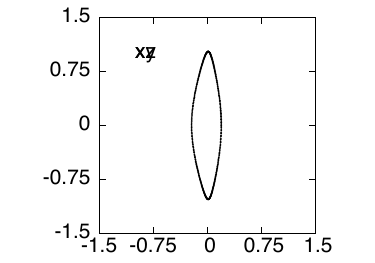}} &
\parbox[c]{2.5cm}{\includegraphics[height=2cm]{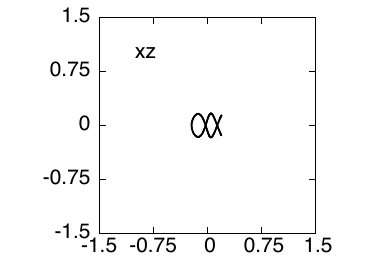}} & 
\parbox[c]{2.5cm}{\includegraphics[height=2cm]{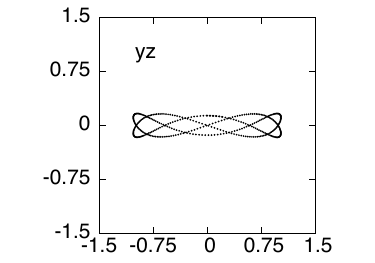}} & 
\parbox[c]{2.cm}{\ej$=-0.316$ \\ $(0.196, 0.135, 0, 0)$. Boxy side-on view.} \\
\hline
\parbox[l]{1.8cm}{\textbf{x1mul2-2u} \\ \textbf{(vm23u)}}& 
\parbox[c]{2.0cm}{-0.3 \\ (2nd tangency of b2 with the $b=-2$ axis).
Initially U. S parts for \ej$ > -0.235$.\\ } & 
\parbox[c]{2.5cm}{\includegraphics[height=2cm]{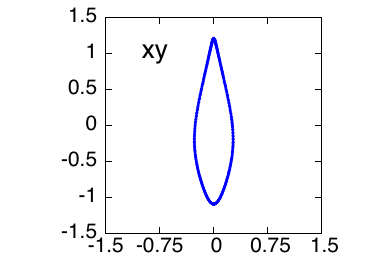}} &
\parbox[c]{2.5cm}{\includegraphics[height=2cm]{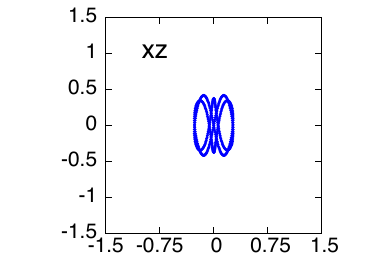}} & 
\parbox[c]{2.5cm}{\includegraphics[height=2cm]{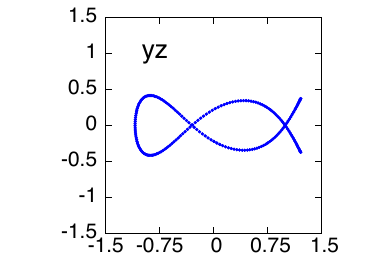}} & 
\parbox[c]{2.cm}{\ej$=-0.316$ \\ $(0.196, 0.135, 0, 0)$.} \\
\hline
\multicolumn{6}{c}{\textbf{x1v3} and \textbf{x1v4} as 2-periodic \citep[see
figures 9 and 10 respectively in][]{spa02a}}\\
\hline
{\textbf{r\_tv1}}& 
\parbox[c]{2.0cm}{$-0.295$ (starting point - inverse bifurcation)\\ (1st intersection
of b1 with the $b=-2$ axis). A group of S, U families with triangular face on 
views and
their radial and vertical bifurcations extending also to lower \ej\!. \\ } & 
\parbox[c]{2.5cm}{\includegraphics[height=2cm]{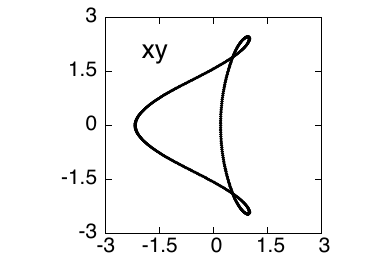}} &
\parbox[c]{2.5cm}{\includegraphics[height=2cm]{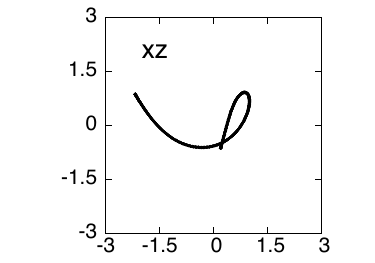}} & 
\parbox[c]{2.5cm}{\includegraphics[height=2cm]{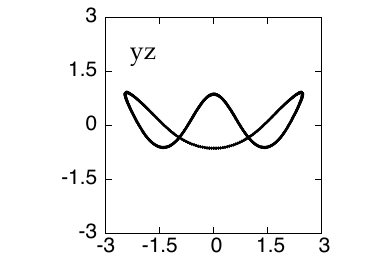}} & 
\parbox[c]{2.cm}{\ej$=-0.2299$ \\ $(0.190, -0.635, 0, 0)$.} \\
\hline
\end{tabular}
\end{table*} 
 
Finally, the 4-periodic orbits are summarized, in the same way as the 2-periodic
ones, in Table~\ref{tab:mul4tab}. We observe that the projections of all
$mul$-periodic
orbits have a morphology that can be described as a Lissajous figure, which becomes
more complicated as the energy, at which a family is bifurcated increases. This is
more evident in the side-on projections of the 3D orbits.

\begin{table*}
\caption{The same as Table~\ref{tab:mul2tab}, now for p.o. of multiplicity 4.}
\label{tab:mul4tab}
\centering
\begin{tabular}{ l  c  c  c  c  c  }
family name & E${_J}{^*}$ & face-on &  end-on & side-on & comments\\
\hline
{\textbf{rm41}}& 
\parbox[c]{2.0cm}{$-0.348$ \\ (1st tangency of b1 with the $b=2$ axis)} & 
\parbox[c]{2.5cm}{\includegraphics[height=2cm]{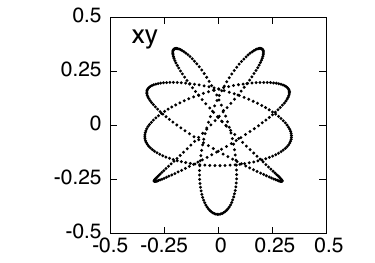}} &
  & 
  & 
\parbox[c]{2.cm}{\ej$=-0.345$ \\ $(0.035, 0, -0.137, 0)$. Similar with face-on
view of orbit in figure 6, 2nd row in GLA.} \\
\hline
{\textbf{vm41}}& 
\parbox[c]{2.0cm}{$-0.341$ \\ (1st tangency of b2 with the $b=2$ axis)} & 
\parbox[c]{2.5cm}{\includegraphics[height=2cm]{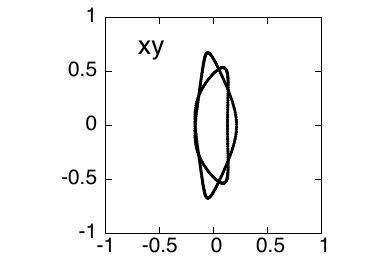}}  &
\parbox[c]{2.5cm}{\includegraphics[height=2cm]{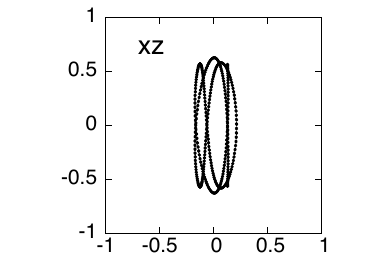}}  & 
\parbox[c]{2.5cm}{\includegraphics[height=2cm]{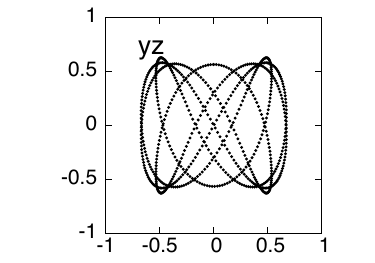}}  & 
\parbox[c]{2.cm}{\ej$=-0.3$ \\ $(0.128, 0.565, 0, 0)$.\\ Face-on similar with figure
10, \mbox{bottom row,} in WM-D. } \\
\hline
{\textbf{vm41u}}& 
\parbox[c]{2.0cm}{$-0.341$ \\ (1st tangency of b2 with the $b=2$ axis)} & 
\parbox[c]{2.5cm}{\includegraphics[height=2cm]{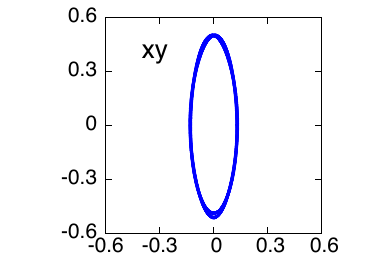}}  &
\parbox[c]{2.5cm}{\includegraphics[height=2cm]{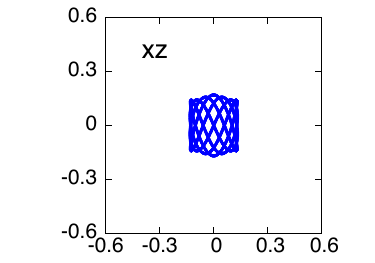}}  & 
\parbox[c]{2.5cm}{\includegraphics[height=2cm]{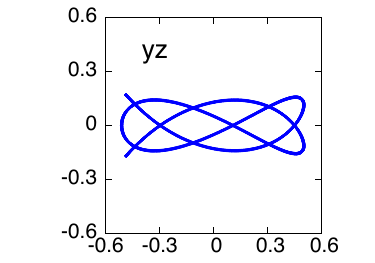}}  & 
\parbox[c]{2.cm}{\ej$=-0.337$ \\ $(0.129, 0.129,$ \\ $-0.0009, 0.038)$.} \\
\hline
{\textbf{vm43}}& 
\parbox[c]{2.0cm}{$-0.326$ \\ (2nd tangency of b2 with the $b=2$ axis)} & 
\parbox[c]{2.5cm}{\includegraphics[height=2cm]{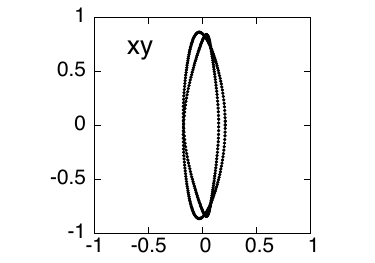}}  &
\parbox[c]{2.5cm}{\includegraphics[height=2cm]{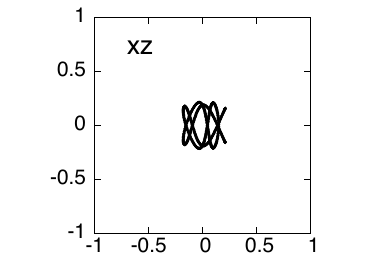}}  & 
\parbox[c]{2.5cm}{\includegraphics[height=2cm]{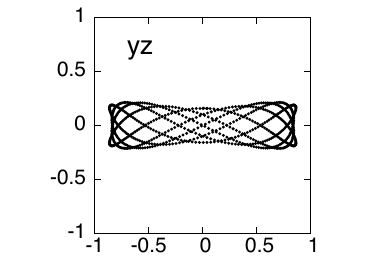}}  & 
\parbox[c]{2.cm}{\ej$=-0.32$ \\ $(0.208 0.158 .0 .0)$.} \\
\hline
{\textbf{rm43}}& 
\parbox[c]{2.0cm}{$-0.317$ \\ (2nd tangency of b1 with the $b=2$ axis)} & 
\parbox[c]{2.5cm}{\includegraphics[height=2cm]{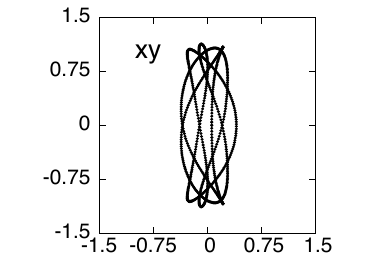}} &
  & 
  & 
\parbox[c]{2.cm}{\ej$=-0.315$ \\ $(0.064, 0, 0, 0)$. Similar with face-on
view of orbit in figure 6, last row, in GLA.} \\
\hline
\end{tabular}
\end{table*}

\section{3-periodic orbits}
\label{sec:3po}
\begin{figure*}
\begin{center}
\resizebox{160mm}{!}{\includegraphics[angle=0]{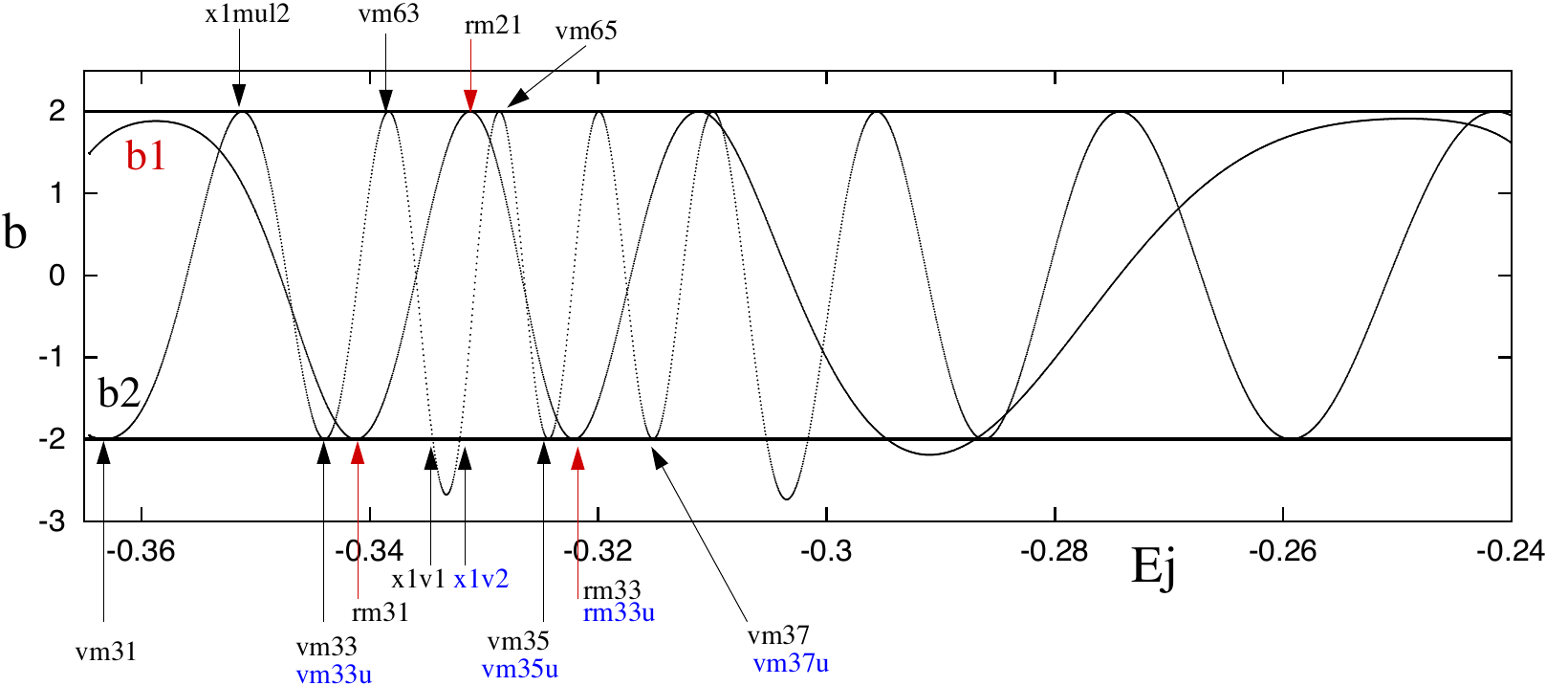}}
\end{center}
\caption{Stability diagram for x1 considered as 3-periodic. Arrows and colors
as in Fig.~\ref{x1stm2}.}
\label{x1stm3} 
\end{figure*}

We followed the same approach for tabulating also the 3- and 6-periodic x1
bifurcations. The corresponding  stability diagram is given in Fig.~\ref{x1stm3}.
Arrows point again to the origin of the families, which are included in
Tables~\ref{tab:mul3tab} and \ref{tab:mul6tab}.

The 3-periodic families also offer some patterns that are frequently encountered
in orbital analyses of barred galaxies models. Among them we find, in a
relatively large range of $\Delta$\ej values, the face-on view of the vm35
family
and the similar, planar, family rm33. This latter family, together with the
2-periodic rm21 one, appear typically on the $(x,p_x)$ Poincar\'{e} surfaces of
section at energies close to the vILR region (e.g. figure 1 in PKb). The
morphological evolution of vm35 gives a nice example of how the evolution of a
projection of a family changes shapes along its characteristic, influenced by 
the
resonances it encounters. 
At the bifurcating point, its face-on view is by default
x1-like. However, at larger energies it develops a face-on morphology
similar to rm33, in the same way as the x1 orbits on the equatorial
plane change their morphology along their characteristic by developing
locally three apocenters at the 3:1 resonance region, four at the 4:1
resonance etc. We can say that the shape of the orbits along a
characteristic of a family ``feels" the resonance it passes by at the
corresponding $\Delta E_J$ regions. Thus, the rm33 face-on pattern
becomes more important for the overall dynamics of the system, as it
can be also encountered at larger energies on the characteristic of one
more, 3D now, family.
The 3D vm35
family in turn, bifurcates at \ej$=-0.3087$ a 6-periodic 3D family that has
edge-on projections with their own morphology and co-exists with the parent
family.
\begin{table*}
\caption{The same as Table~\ref{tab:mul2tab}, now for p.o. of multiplicity 3.}
\label{tab:mul3tab}
\centering
\begin{tabular}{ l  c  c  c  c  c  }
family name & E${_J}{^*}$ & face-on &  end-on & side-on & comments\\
\hline
{\textbf{vm31}}& 
\parbox[c]{2.0cm}{$-0.363$ \\ (1st tangency of b2 with the $b=-2$ axis)} & 
\parbox[c]{2.5cm}{\includegraphics[height=2cm]{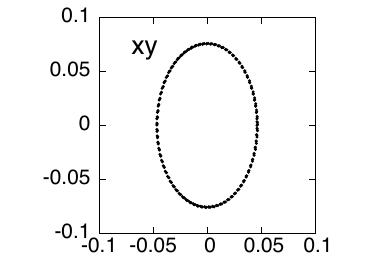}}  &
\parbox[c]{2.5cm}{\includegraphics[height=2cm]{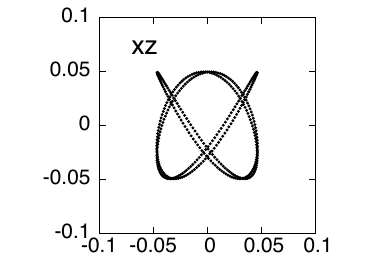}}  & 
\parbox[c]{2.5cm}{\includegraphics[height=2cm]{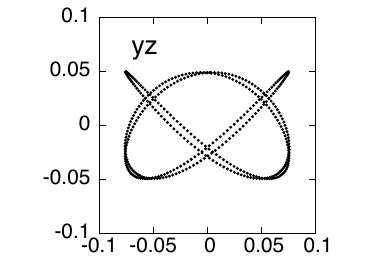}}  & 
\parbox[c]{2.cm}{\ej$=-0.362$ \\ $(0.046, 0.049,$ \\ $-0.0007, -0.003)$.
\mbox{}} \\
\hline
{\textbf{vm33}}& 
\parbox[c]{2.0cm}{$-0.344$ \\ (2nd tangency of b2 with the $b=-2$ axis)} & 
\parbox[c]{2.5cm}{\includegraphics[height=2cm]{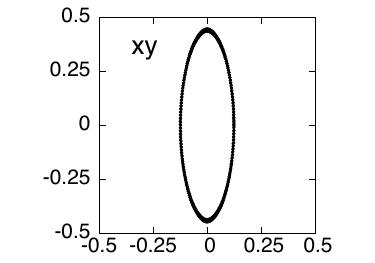}}  &
\parbox[c]{2.5cm}{\includegraphics[height=2cm]{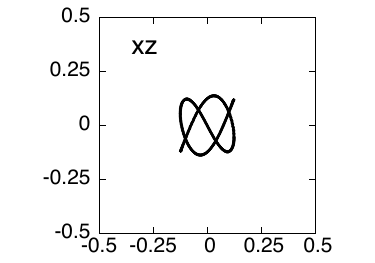}}  & 
\parbox[c]{2.5cm}{\includegraphics[height=2cm]{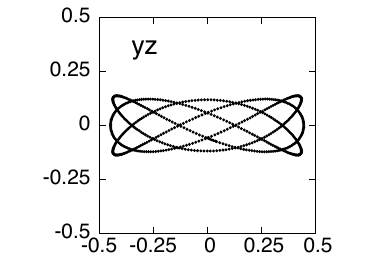}}  & 
\parbox[c]{2.cm}{\ej$=-0.341$ \\ $(0.124, -0.058,$ \\ $-0.001, -0.073)$.} \\
\hline
{\textbf{vm33u}}& 
\parbox[c]{2.0cm}{$-0.344$ \\ (2nd tangency of b2 with the $b=-2$ axis)} & 
\parbox[c]{2.5cm}{\includegraphics[height=2cm]{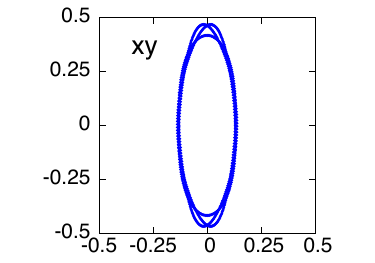}}  &
\parbox[c]{2.5cm}{\includegraphics[height=2cm]{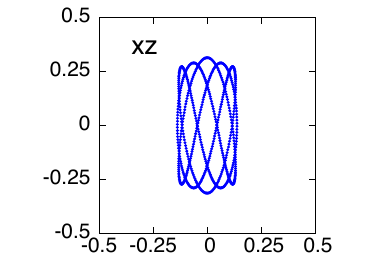}}  & 
\parbox[c]{2.5cm}{\includegraphics[height=2cm]{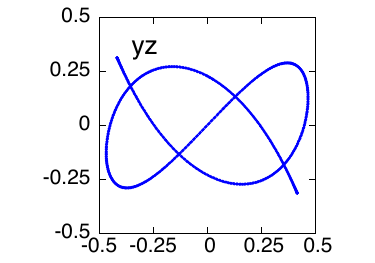}}  & 
\parbox[c]{2.cm}{\ej$=-0.330$ \\ $(0.130, -0.229,$ \\ $0.006, -0.091)$.
\\ \mbox{side-on view} \mbox{similar with} \mbox{figure 6, last} row in AVSD,
\mbox{figure 2, 3rd row} \mbox{and figure 5 down} in PWG,\\ \mbox{figure 4 last
row} in VSAD, figure 21 2nd and 3rd columns in WAM.} \\
\hline
{\textbf{rm31}}& 
\parbox[c]{2.0cm}{$-0.341$ \\ (1st tangency of b1 with the $b=2$ axis)} & 
\parbox[c]{2.5cm}{\includegraphics[height=2cm]{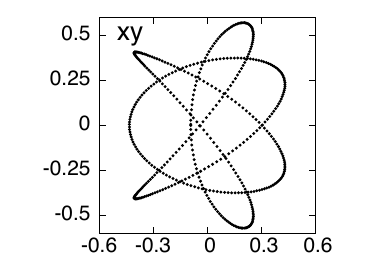}} &
  & 
  & 
\parbox[c]{2.cm}{\ej$=-0.333$ \\ $(-0.091, 0, 0, 0)$.} \\
\hline
\multicolumn{6}{c}{\textbf{x1v1} and \textbf{x1v2} as 3-periodic}\\
\hline
{\textbf{vm35}}& 
\parbox[c]{2.0cm}{$-0.324$ \\ (3rd tangency of b2 with the $b=-2$ axis)\\ S: \ej
$< -0.304$.} & 
\parbox[c]{2.5cm}{\includegraphics[height=2cm]{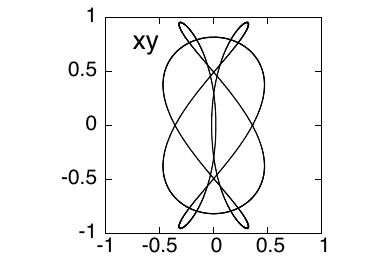}}  &
\parbox[c]{2.5cm}{\includegraphics[height=2cm]{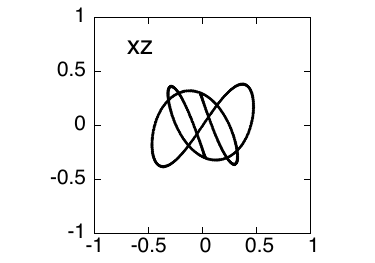}}  & 
\parbox[c]{2.5cm}{\includegraphics[height=2cm]{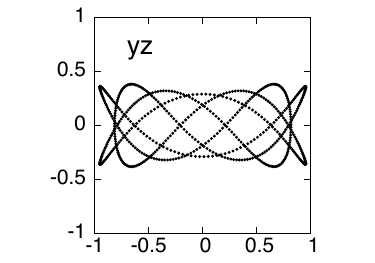}}  & 
\parbox[c]{2.cm}{\ej$=-0.305$ \\ $(0.358, -0.187,$ \\ $0.105, 0.143)$.
\\Face-on similar with figure 4 in PKb and with figure 7, last row, in GLA.} \\
\hline
{\textbf{vm35u}}& 
\parbox[c]{2.0cm}{$-0.324$ \\ (3rd tangency of b2 with the $b=-2$ axis)\\
Initially (and mostly) U, Small S and $\Delta$ parts close to \ej
$\lessapprox -0.28$} & 
\parbox[c]{2.5cm}{\includegraphics[height=2cm]{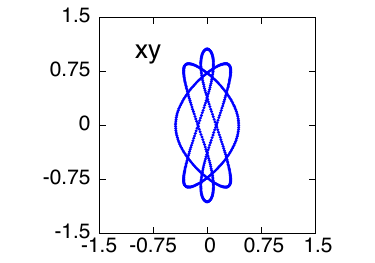}}  &
\parbox[c]{2.5cm}{\includegraphics[height=2cm]{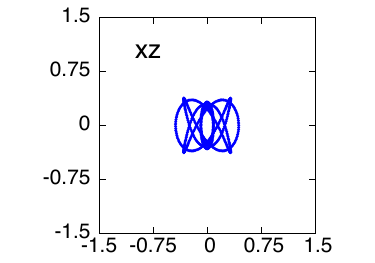}}  & 
\parbox[c]{2.5cm}{\includegraphics[height=2cm]{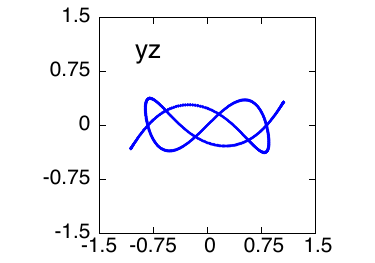}}  & 
\parbox[c]{2.cm}{\ej$=-0.307$ \\ $(0.127, -0.239,$ \\ $0.087, -0.096)$.
\\ } \\
\hline
{\textbf{rm33}}& 
\parbox[c]{2.0cm}{$-0.322$ \\ (2nd tangency of b1 with the $b=2$ axis)} & 
\parbox[c]{2.5cm}{\includegraphics[height=2cm]{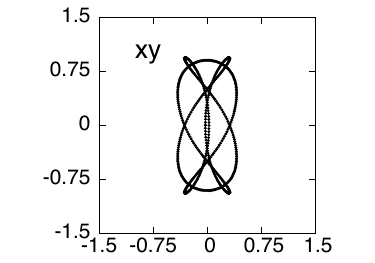}} &
  & 
  & 
\parbox[c]{2.cm}{\ej$=-0.318$ \\ $(-0.0257, 0, 0, 0)$.\\ Same with orbit in
figure 4 in PKb, similar with orbit in figure 7, last row, in GLA.} \\
\hline
\end{tabular}
\end{table*} 

\begin{table*}
\contcaption{ }
\centering
\begin{tabular}{ l  c  c  c  c  c  }
\hline
{\textbf{rm33u}}& 
\parbox[c]{2.0cm}{$-0.322$ \\ (2nd tangency of b1 with the $b=2$ axis)} & 
\parbox[c]{2.5cm}{\includegraphics[height=2cm]{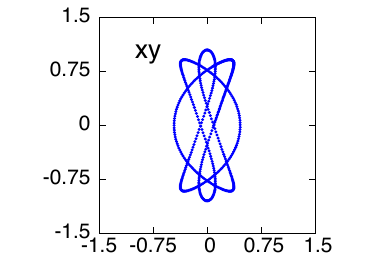}} &
  & 
  & 
\parbox[c]{2.cm}{\ej$=-0.317$ \\ $(0.0916, 0, 0, 0)$.} \\
\hline
{\textbf{vm37}}& 
\parbox[c]{2.0cm}{$-0.315$ \\ (4th tangency of b2 with the $b=-2$ axis)\\ S in
general.} & 
\parbox[c]{2.5cm}{\includegraphics[height=2cm]{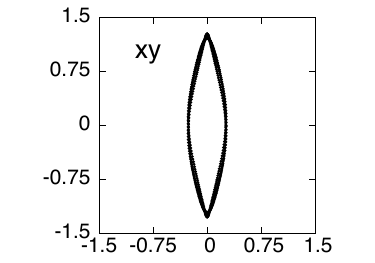}}  &
\parbox[c]{2.5cm}{\includegraphics[height=2cm]{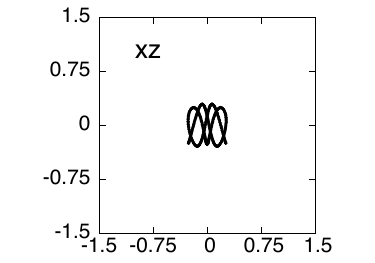}}  & 
\parbox[c]{2.5cm}{\includegraphics[height=2cm]{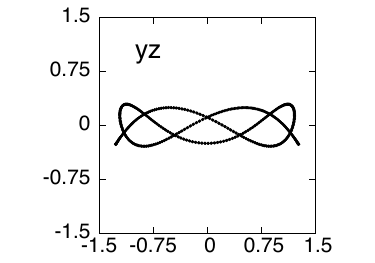}}  & 
\parbox[c]{2.cm}{\ej$=-0.304$ \\ $(0.262, 0.112,$ \\ $-0.012, -0.127)$.
\\} \\
\hline
{\textbf{vm37u}}& 
\parbox[c]{2.0cm}{$-0.315$ \\ (4th tangency of b2 with the $b=-2$ axis)\\ U in
general.} & 
\parbox[c]{2.5cm}{\includegraphics[height=2cm]{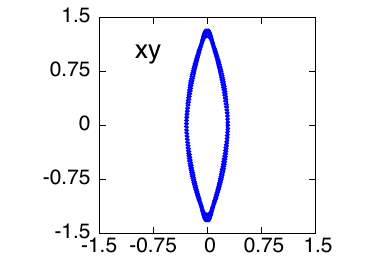}}  &
\parbox[c]{2.5cm}{\includegraphics[height=2cm]{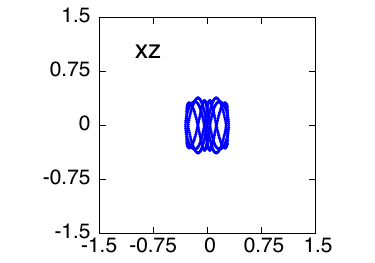}}  & 
\parbox[c]{2.5cm}{\includegraphics[height=2cm]{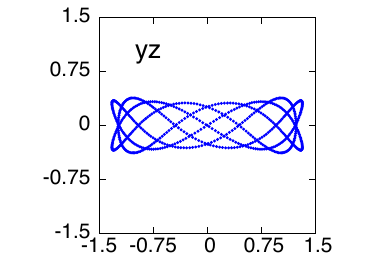}}  & 
\parbox[c]{2.cm}{\ej$=-0.298$ \\ $(0.278, -0.258,$ \\ $-0.157, -0.096)$.
\\} \\
\hline
& & & & & \\
\end{tabular}
\end{table*}

\begin{table*}
\caption{The same as Table~\ref{tab:mul2tab}, now for p.o. of multiplicity 6.}
\label{tab:mul6tab}
\centering
\begin{tabular}{ l  c  c  c  c  c  }
family name & E${_J}{^*}$ & face-on &  end-on & side-on & comments\\
\hline
\multicolumn{6}{c}{\textbf{x1mul2} as 6-periodic}\\
\hline
{\textbf{vm63}}& 
\parbox[c]{2.0cm}{$-0.338$ \\ (2nd tangency of b2 with the $b=2$ axis)} & 
\parbox[c]{2.5cm}{\includegraphics[height=2cm]{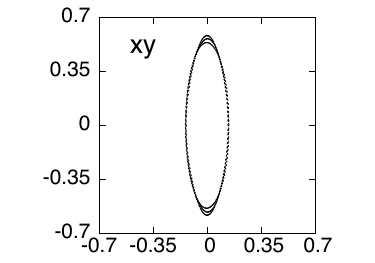}}  &
\parbox[c]{2.5cm}{\includegraphics[height=2cm]{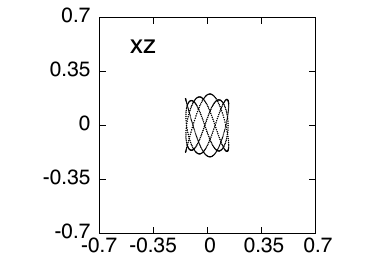}}  & 
\parbox[c]{2.5cm}{\includegraphics[height=2cm]{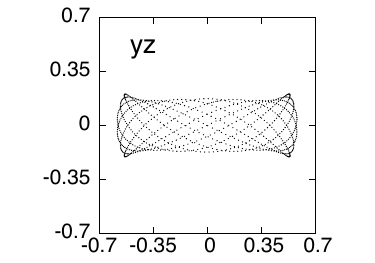}}  & 
\parbox[c]{2.cm}{\ej$=-0.333$ \\ $(0.131,0.140,\\
-0.0035,0.061)$.\\ Similar to x1 quasi-periodic orbits, figure 6 in PKa.  } \\
\hline
\multicolumn{6}{c}{\textbf{rm21/rm22} as 6-periodic}\\
\hline
{\textbf{vm65}}& 
\parbox[c]{2.0cm}{$-0.328$ \\ (3rd tangency of b2 with the $b=2$ axis)} & 
\parbox[c]{2.5cm}{\includegraphics[height=2cm]{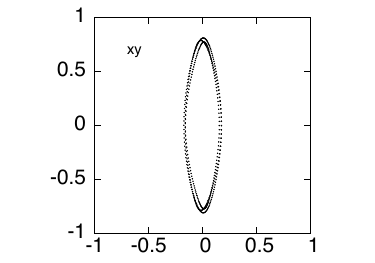}}  &
\parbox[c]{2.5cm}{\includegraphics[height=2cm]{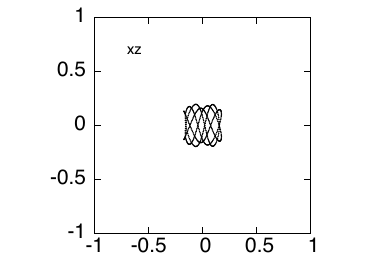}}  & 
\parbox[c]{2.5cm}{\includegraphics[height=2cm]{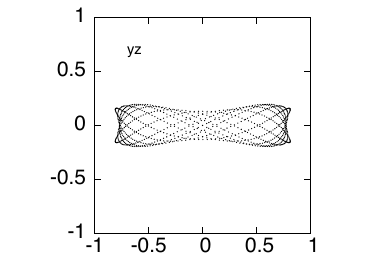}}  & 
\parbox[c]{2.cm}{\ej$=-0.24$ \\ $(0.173,0.105,\\
0.001,0.066)$. \\ Similar to x1 quasi-periodic orbits, figure 15 in PKa.} \\
\hline
\end{tabular}
\end{table*} 

In Table~\ref{tab:mul6tab}, where we present 6-periodic orbits, we meet again
families
that we have already presented as 2-periodic, at tangencies of the 3-periodic x1
stability curves with the $b=2$ axis. Thus, the first to appear in the table is the
known x1mul2 family, while at a larger energy in the stability diagram of
Fig.~\ref{x1stm3}, we meet again rm21 and rm22.

For the 3- and 6-periodic orbits, we observe again that the larger the energy,
the more complicated, more ``dense'', Lissajous figures we have in their edge-on
profiles. In most cases of the 3D families in Tables~\ref{tab:mul3tab} and
\ref{tab:mul6tab} we have more distinguishable shapes in the side-on profiles of
the unstable periodic orbits and on the end-on of the stable ones (a kind of
exception seems to be the vm37 - vm37u pair). This means that it is easier for
the side-on profiles based on 3- and 6-periodic stable orbits to obtain a
compact, boxy, character.

As multiplicity and energies increase, we find profiles of p.o. that resemble
more those of 3D quasi-periodic orbits trapped around x1. Characteristic
examples are the side-on projections of the two 6-periodic orbits we present in
Table~\ref{tab:mul6tab}, as well as the one of the p.o. vm43
(Table~\ref{tab:mul4tab}). The local minimum of the $z$ extent at $y$=0 in the
side-on views of vm65 and vm43 is reminiscent of the corresponding minimum of
the x1 quasi-periodic orbits on invariant tori approaching the x1v2 initial
conditions (see figure 15 in PKa), while the side-on view of vm63, having a
local very shallow maximum at (0,0),  points to side-on profiles of 3D
quasi-periodic orbits of x1 approaching the energy at which x1v1 is bifurcated
(figure 2 in PKa). Such orbits have to be counted among those that reinforce
peanut-shaped morphologies.

For the sake of completeness we have calculated the families of 5-periodic
orbits as well. We do not present them here separately, since their edge-on
morphologies can be simply described as ``boxy'', complicated Lissajous figures.
They are found combined with any kind of face-on morphologies presented in the
tables, from x1 ellipses to rather ``filled'' shapes (cf for example the face on
views of vm41u or vm43 with rm43 in Table~\ref{tab:mul4tab}).

\section{multi-periodic, quasi- and non-periodic orbits at high energies}
\label{sec:highej}
Bifurcations introduce complexity in a dynamical system. In our rotating bar 
model the families bifurcated from the x1 family have more complicated 
morphologies than those of the x1 p.o., because they develop asymmetries or 
loops. This results generally to shapes that are less elongated than those of 
the x1 ellipses at the same energy (e.g. the 3:1-type bifurcations). This is 
particularly the case for the shapes of the $mul$-periodic orbits, whose
outline 
become evidently ``rounder'' than the x1 already at energies close to the one at 
which they have been bifurcated. The effect is more conspicuous in the 
projections of the 3D bifurcations of x1 on the equatorial plane.

\begin{figure}
\begin{center}
\resizebox{80mm}{!}{\includegraphics[angle=0]{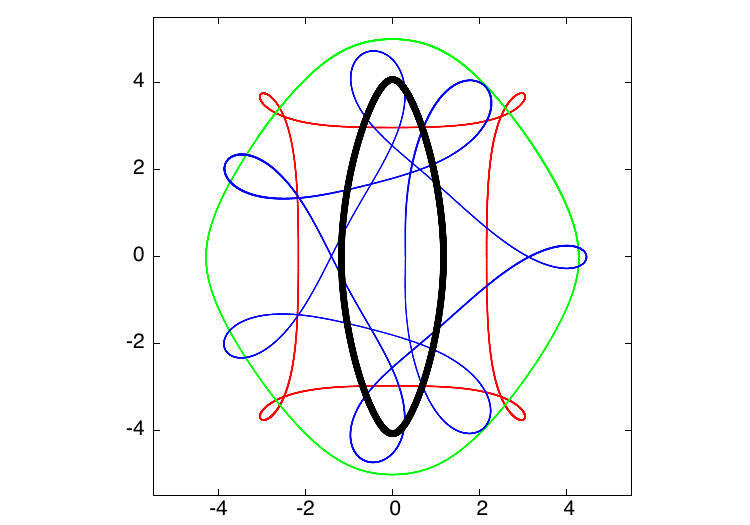}}
\end{center}
\caption{Typical periodic orbits at the outer parts of the model. In black we 
give
one of the longest x1 p.o. at \ej $= -0.22$, while the three others (red, blue and
green) are at \ej $= -0.2$. The blue, 3-periodic orbits is typical of the shape
of $mul$-periodic orbits we encounter at high energies. Only the x1 p.o. can be
considered as bar-supporting.}
\label{heorbs2d} 
\end{figure}

In the particular Ferrers bar model we use in this work, the x1
characteristic and stability curves do not have a smooth evolution at high
energies. They are characterized by foldings and ``bows'' respectively
\citep{spa02a}. At any rate, the longest x1 p.o. reaches a $y \approx 4.1$
value, which can be considered as an estimation of the maximum radius of the
bar we can build with the orbital content of this model. At high
energies, the ``round'' shapes of the x1 bifurcations, and especially those of
the $mul$-periodic ones, spread out beyond the extent of the x1 in the
direction 
of the bar minor axis.  Thus they
can not be part of the bar. Since they are not elongated along the major axis of
the bar, the only way to contribute to its density would be to build
substructures \textit{within} the bar. We give in Fig.~\ref{heorbs2d} typical
p.o. that dominate the outer parts of the model and are suitable for describing
the situation we refer to. The black ellipse is the x1 p.o. for \ej $= -0.22$,
close to the longest x1 in the system, while the other three plotted are for \ej
$=-0.2$. The rectangular-like x1 p.o. is the x1 representative at this energy
(plotted with red). It is stable and already considerably square, so it does not
help the bar to extend to larger distances from the center. In other models,
chaotic orbits, sticky to rectangular-like periodic ones, could support an outer
boxiness of the bar \citep{paq97}. In this model such orbits exist in a small
$\Delta$\ej interval. In any case, this refers essentially to the shape of the
outermost x1 p.o. and not to the presence of another orbit, let alone of a
$mul$-periodic bifurcation. The rhomboidal, green p.o., typical at the 4:1
resonance of rotating galactic models, could be associated only with inner rings
or perhaps lenses. The blue, 3-periodic orbit at the same energy, could only
contribute to the enhancement of a disc surrounding the bar. Evidently, such
orbits could not be used as a basis for supporting thick bar structures either.
For this reason we do not proceed to a detailed presentation of orbits that are
obtained when the radial bifurcations of x1 at high \ej are
vertically perturbed.

As regards the 3D families of the x1-tree, which support the bar, they
provide building blocks that stay closer and closer to the equatorial plane as
energy increases \citep{psa02}. They are organized in blocks belonging
to stable
x1v$n$ families that reach a maximum radius, $R_{max}$, on the equatorial plane,
beyond which  only the height of their individual orbits increases. This maximum
radius can be considered as a maximum distance within which these orbits can
support the bar. 
In this way the edge-on view forms a profile with the outline of a 
staircase
(we refer to it as a ``stair-type'' profile) and
in which the larger the $n$ of the parent x1v$n$ family, the more vertically
thin, within $R_{max}$, the building block will be \citep[cf figures 11 to 19
in][]{psa02}. The shapes of their 
radial bifurcated orbits become "rounder" as \ej
increases, as already noted for the orbits radially bifurcated from x1. 
However, apart from the round shapes that render them as
non-bar supporting, the radial bifurcations of the x1v$n$ families in the model
have parts of their characteristic that extend towards the center of the system,
bringing members of these families to the central regions. However, none of the
encountered morphologies had the dimensions to characterize a face-on or edge-on
profile. 

\begin{figure}
\begin{center}
\resizebox{80mm}{!}{\includegraphics[angle=0]{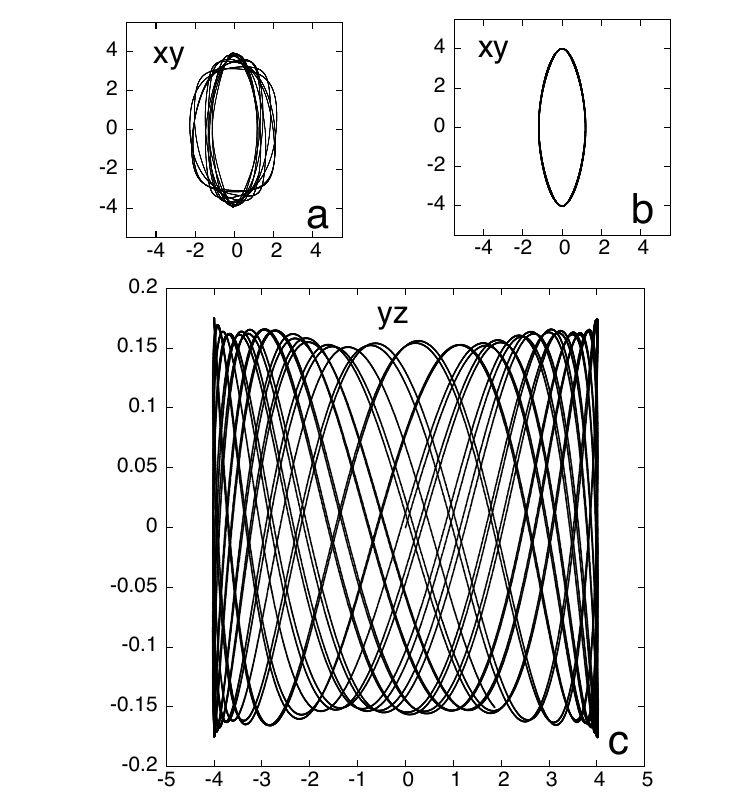}}
\end{center}
\caption{Verically perturbed x1 orbits at \ej$=-0.22$. In (a) is given the face-on
projection of the orbit perturbed by $p_z=0.1$, while in (b) by $p_z=0.05$. In (c)
we give the side-on view of the orbit in (b), with different scale in the axes. This
is a profile typical for vertically perturbed x1 orbits.}
\label{hex1z} 
\end{figure}

As the energy increases, the structure of the phase space of a rotating 3D bar
becomes increasingly complex, since the number of existing families of p.o.
increases and their stability varies. Chaos increases not only because the
volume of the chaotic seas becomes larger, but also because small deviations
from the initial conditions of an orbit may bring it on different zones of
influence of the numerous p.o. existing at the same energy. In other words the
shape of a structure that will be possibly supported is sensitive to the
perturbation that we apply. This dependence is clearly non-monotonic. It is
technically not easily workable to isolate these zones of influence around the
p.o. in the 4D space of section of a 3D system. While we have a fair
understanding of the contribution of the 3D families of the x1 tree to the
backbone of 3D bars \citep{psa02} and knowledge of the contribution of quasi-
and non-periodic orbits at the ILR regions of the models \citep[][PKa,
PKb]{kpc13}, less work has been done in assessing the role of 3D quasi- and
non-periodic orbits around the planar x1 orbits \citep{cppsm17}. Here we give in
Fig.~\ref{hex1z} an example that delineates how a vertically perturbed x1 orbit
close to the end of the bar could contribute to the edge-on profile of our
model. We consider the x1 orbit at \ej$=-0.22$, which we see in
Fig.~\ref{heorbs2d} and we start increasing $p_z$. The face-on view of the orbit
when perturbed by $p_z=0.1$ fills a thick ring (Fig.~\ref{hex1z}a), similar to
quasi-periodic orbits obtained when the orbit is perturbed radially. The orbit
is on a torus, i.e. viewed edge-on, it has effectively rectangular-like edge-on
projections that can be vaguely described as complicated Lissajous figures
reaching a height about 0.4 above the equatorial plane. Similar edge-on
profiles are obtained for smaller perturbations as well, however the height they
reach is much smaller. In Fig.~\ref{hex1z}b we give the face-on projection of
the x1 orbit perturbed by $p_z=0.05$, which is almost identical to the 
planar x1
orbit. However, its side-on view given in Fig.~\ref{hex1z}c is again rectangular
like. The orbits have been integrated for seven x1 periods. We did not find at
large energies any simple $mul$-periodic edge-on morphology like those
encountered
in the central parts of the model and described in the tables.

\begin{figure}
\begin{center}
\resizebox{80mm}{!}{\includegraphics[angle=0]{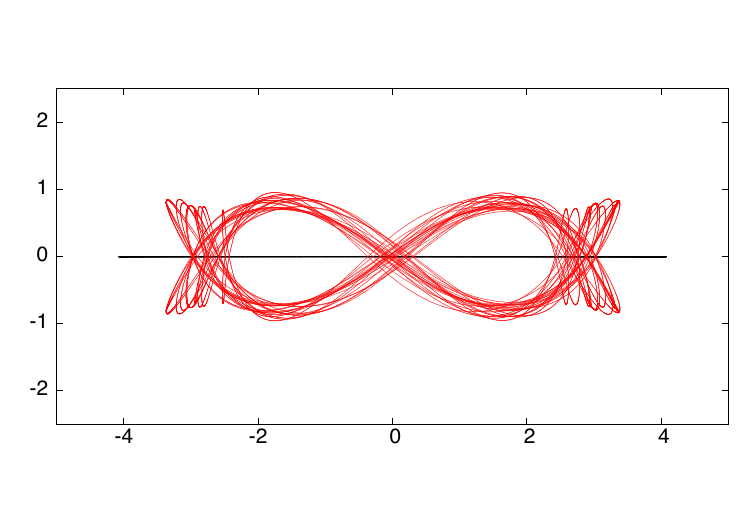}}
\end{center}
\caption{The side-on projection of a x1 orbit at
\ej$=-0.22$, perturbed by $p_z=0.2$ that reaches a x1v4 torus. The black horizontal
line, indicates the length of x1 at the same energy.}
\label{x1v4} 
\end{figure}

By increasing the vertical perturbations even more we do not necessarily
enter a
chaotic sea. Moving along a direction in phase space we may reach tori
of another family. By reaching a perturbation $p_z=0.2$ of the x1 orbit of
Fig.~\ref{heorbs2d}, i.e. by applying a larger vertical perturbation than in the
case of the orbits of Fig.~\ref{hex1z}, we reach a quasi-periodic orbit around x1v4,
which for \ej$=-0.22$ is stable in our model. Such orbits, if populated, will
support a peanut shaped thick bar in which the thick part is almost the bar itself
and not part of it as we can observe in Fig.~\ref{x1v4} \citep[see also figure 1c
in][]{psa02}. 
In principle, in our model we can find edge-on profiles that are determined by
simple periodic orbits of the x1 tree at high energies. However, $mul$-periodic,
vertical, 3D bifurcations of the x1v$n$ families with $n\geq 3$,  are rather
insignificant. In most cases they exist only in narrow $\Delta$\ej intervals and
their characteristics fall again on the characteristic of the parent family.
Thus, at high energies the profiles that dominate are mainly those of
complicated Lissajous figures or one of the known shapes of the
\textit{simple}-periodic orbits of the x1-tree. 

We conclude that the role of $mul$-periodic orbits is reduced as energy
increases.
Practically, the shapes of the most important $mul$-periodic orbits can play a
role only in the thick part of the bar, which in the case we study here, as in
most cases of barred galaxy models, ends between 0.3 and 0.7 times the length of 
the bar \citep[e.g. Fig. 11 in][]{a15}.

The reader interested in the comparison of the orbital shapes published
in the papers mentioned in the very beginning of the introduction with the 
morphology of the periodic orbits presented in the tables, may proceed at this
point to Appendix~\ref{sec:qporbs}. In the next section we discuss the main
results of our study.

\section{Discussion and Conclusions}
\label{sec:concl} 
We studied here the origin of morphologies appearing in the projections of
orbits of multiplicity higher than one and which are not trapped in the
immediate neighborhood of the x1 family or of the families of the x1 tree. We
investigated the possibility of having bars built with orbits that deviate
considerably from elliptical-like morphologies. 
Our study was motivated by the existence of such orbits, mentioned lately in a
number of papers about the orbital content of 3D galactic bars. 

The present work is
mainly comparative and aims to the understanding of the origin of such orbits
and of their shapes, a study sorely missing from all papers
presenting and discussing their morphologies. The vast majority of the orbits
found in published papers are quasi- or non-periodic. Thus, they can be either
trapped on invariant tori around a stable periodic orbit, or they will be
diffusing in a chaotic sea. A kind of intermediate situation are the sticky
chaotic orbits \citep[see e.g.][]{ch08}, in which an ultimately chaotic orbit
behaves for a relatively long time as regular, supporting a particular
structure. Below, we summarize the basic subjects discussed through our
paper and the corresponding main results we reached in each case:

\subsection{3D Quasi- and non-periodic orbits may combine in their three
projections morphologies from projections of different families of p.o:}

The morphology of a 3D orbit exhibits in its three projections in general
patterns that have been introduced in radial or vertical resonances, at lower
energies than the energy of the orbit. This can happen in two ways. Firstly,
when a 3D family is bifurcated from a parent one as \ej increases, it may retain
either the edge-on or the face-on morphology of the parent family, depending on
whether the radial or the vertical stability index reaches a $b=\pm2$ axis,
respectively. A new family will have in its projections a combination of old and
new features. Secondly, at a certain \ej, if we start perturbing the initial
conditions of a periodic orbit in phase space we will approach those of another
p.o. family. By doing so, we observe a smooth, gradual transition from a
certain morphology of a quasi- or non-periodic orbit to another morphology. The
navigation in the 4D space of section of an autonomous Hamiltonian system has
several practical complications. Nevertheless, an example of the smooth
variation of the morphology of the orbits as we depart from the initial
conditions of a periodic one has been given in PKa. Moving along the $p_z$
direction from x1 towards x1v2, the 3D quasi-periodic orbits of the stable x1
orbit start resembling in their side-on profiles the unstable x1v2 orbit. When
we cross the last invariant torus around x1 and we enter the 4D chaotic sea, the
unstable manifold of the x1v2 leads us around the x1v1 tori and then the side-on
projections of the orbits exhibit hybrid x1v1-x1v2 morphologies \citep[see
figure 12 in][and figure 13 in PKa]{kpc13}. Such morphological transformations
are typical as one moves in the phase space, although we have not yet
established rules to predict them. For the needs of the present paper we
keep in mind that the individual morphological patterns presented in the 
columns of
Tables~\ref{tab:mul2tab} to \ref{tab:mul6tab} can be encountered also combined 
in the
projections of quasi- and non-periodic orbits in several models.

\subsection{About the information included in the Tables with the
$mul$-periodic orbits:}

Since the detailed discussion of all families bifurcating from the x1-tree and
of their evolution would be impractical, we presented the main orbital patterns
summarized in Tables. In these Tables we refer to the origin of 
the
various morphological patterns published in several papers. A secure conclusion
is that they have their origin in the second type bifurcations of x1. The best
way of presenting how they are linked to x1 is to find the tangencies (or
intersections) of the stability curves of x1 with the $b=2$ and $b=-2$ axes when
it is considered $mul$-periodic. Since x1 is almost everywhere stable, at the
tangencies, there are always bifurcated pairs of one initially stable and one 
initially
unstable family. Because of the symmetries in our model, these orbits have
also symmetric counterparts with respect to axes of the system. These ``twin''
orbits share the same stability curves. In practice, the p.o. presented
in the tables of the present paper, together with the orbits of the x1-tree in
\citet{psa02}, can be considered as the basis for most orbital patterns expected
to exist in many standard 3D N-body barred galaxy models.

\subsection{Where do we find each orbital morphology on the spaces of
section?}

By comparing the shapes of the orbits in our Tables with the structures of the
published orbits in the models, it is obvious that not all orbital patterns are
encountered with the same frequency. This may be due to the fact that the second
type bifurcations we study, occupy parts of the stability islands centred on the
x1 p.o., as we can realize in the 2D case of orbits on the equatorial plane (see
e.g. Fig.~\ref{ofmainm2}b). Such islands have always central parts with
invariant curves belonging to quasi-periodic orbits morphologically influenced
by the shape of the central p.o. In the case of x1, they are elliptical-like.
Then, at larger distances from the center of the island, we have, as
expected, a ring of smaller islands surrounding the central part. In the case
described in Fig.~\ref{ofmainm2}b the region of this ring is under the rm21/rm22
and rm21u/rm22u morphological influence. The area of phase space where the 
initial conditions of a quasi- or non-periodic orbit with a given morphology 
will be located is determined by the degree of perturbation of the p.o.


In 3D cases it is more difficult to isolate the areas of morphological
influence. It is characteristic that in all cited studies the frown - smiles and
``$\infty$''-like profiles are included (in AVSD figure 6, 3rd row;
\citet{cppsm17} figure 9, b4, figure 10, b2 and b3; GLA all orbits in figure 7;
PKa especially figures 11 and 15; PWG figure 2, panels E and F; VSAD figure 4,
4th and 5th rows; WAM figure 5). Both x1v1 and x1v2 side-on profiles are the
smoking gun of the existence of the vILR in the system, being introduced
together at this resonance. Recently, \citet{ph18} have shown that it is the
presence of the vILR that offers building blocks for the peanuts and as such can
serve either regular or sticky chaotic orbits. Thus, the presence of the
resonance is more important than the stability of the orbits. In PKa and in
\citet{ph18} it is shown that the shapes of several 3D quasi-periodic orbits
around x1, have a morphology similar to that of the unstable one of the x1v1,
x1v2 pair. Thus, in general, in the presence of a vILR resonance in a
system, in order to find initial conditions of orbits that do not follow
frown-smiles-like, or $\infty$-like trajectories in their side-on projections
(at least during an important time interval), we have to avoid considering
initial conditions in the regions of phase space, which are influenced by these
three families. 
For this purpose, and as indicated by the figures in
Katsanikas et al. 2013, PKa and Patsis \& Harsoula (2018), we
have to stay away from large volumes of phase-space
around the main families of periodic orbits encountered beyond the
vILR energy (in models that experience this resonance).

%


\subsection{Limits in the $\Delta$\ej influence of a family, imposed by the
evolution of its characteristic curve:}

\begin{figure}
\begin{center}
\resizebox{85mm}{!}{\includegraphics[angle=0]{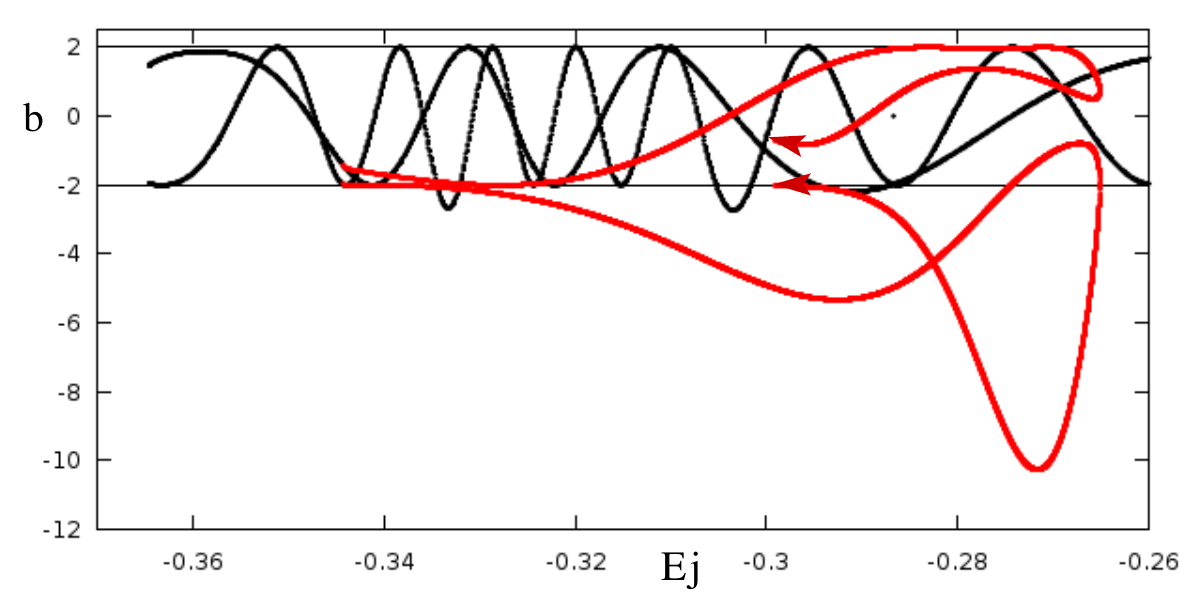}}
\end{center}
\caption{The evolution of the stability indices of the vm33u family (red
curves),
bifurcating from those of x1 at \ej $\approx -0.344$. After reaching a maximum
\ej
the vm33u family continues towards smaller energies. }
\label{stabue} 
\end{figure}

An ultimate limit of the extent of a family is put by its maximum \ej value
beyond which the family does not exist. In such cases the characteristic of the
family reaches a maximum \ej and then turns towards smaller \ej values. The most
straightforward example one can bring is the x1 family itself in the case of
type II gaps at the radial 4:1 resonance (Contopoulos 1988). In such a case,
following the characteristic curve, we find that, as \ej increases, the $x$ 
value
first increases to reach a maximum $x_0$ value and then decreases until a
maximum \ej value. At this point it turns backwards towards smaller \ej values
and towards the centre of the system \citep[see e.g. figure 1 in][]{spa02a}.
Such regressions of the characteristic are especially important for the dynamics
of slowly rotating bars \citep{spa02b, tp15}. They are typical in the evolution
of the families we presented in the tables. As an example we present in
Fig.~\ref{stabue} the evolution of the 
stability indices of the vm33u family (red curves), after bifurcating from x1
considered 3-periodic (black curves). The morphology of this family is
encountered in works by AVSD, PWG, VSAD and WAM. In this case
$E_{J_{max}}\approx -0.265$. This evolution of the characteristic is common
among the $mul$-periodic orbits we found. One could conclude that the backwards
evolving branch of the characteristic will bring more representatives of the
family in the phase space at lower energies and thus its role in influencing
larger areas of the phase space would be more important. 

\subsection{Different families of p.o. offer different sizes of peanut
building blocks.}

In the case of the peanut structure, apart from the evolution of the
characteristic of a family, there is also another property that has to be taken
into account when its orbits are used as a building block for it. It is the
length along the bar's major axis within which this family can support the
peanut and in general the size of the peanut it could support.

\begin{figure}
\begin{center}
\resizebox{85mm}{!}{\includegraphics[angle=0]{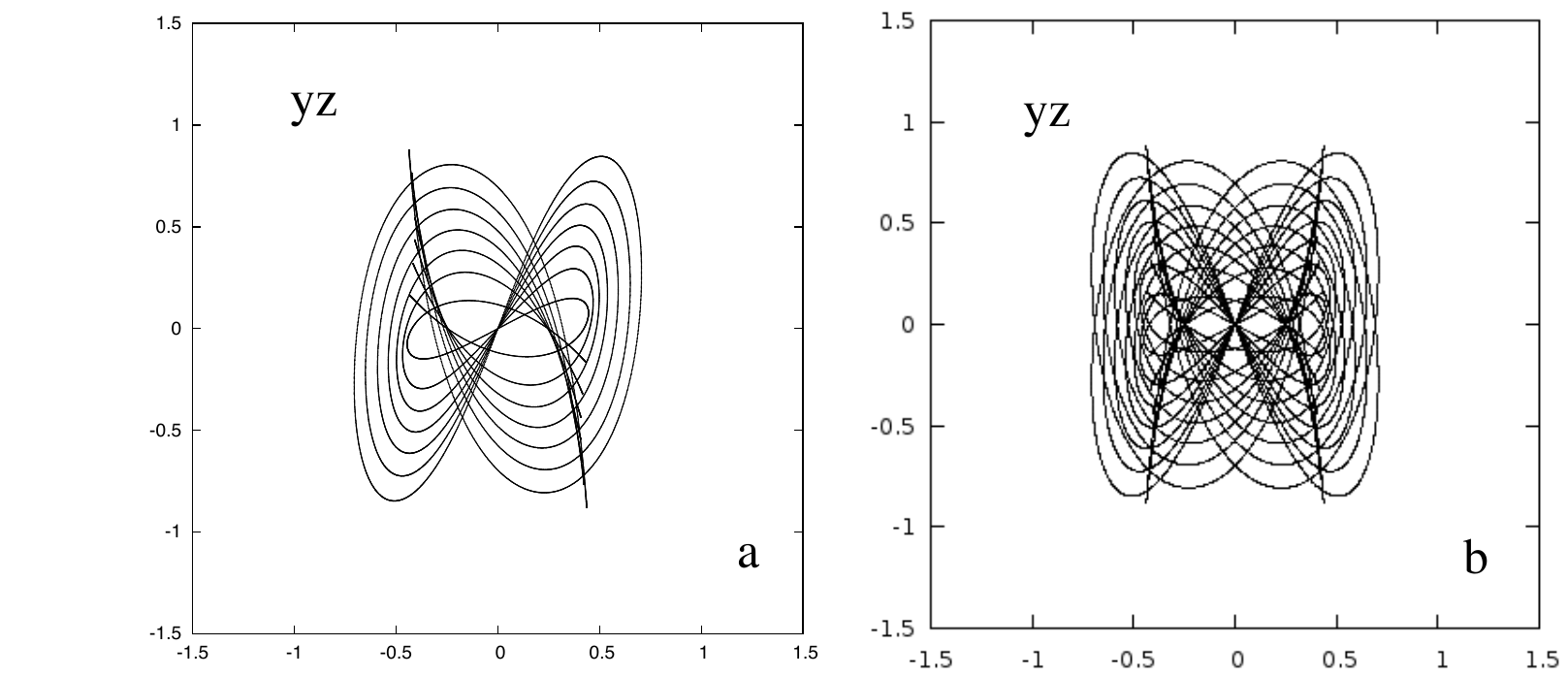}}
\end{center}
\caption{Seven vm33u periodic orbits at energies $-0.34\leq$ \ej
$\leq -0.28$. In (a) we consider orbits belonging to one branch of the family,
while in (b) we consider orbits from both branches symmetric with respect to the
equatorial plane. In (b) we observe the appearance of two X
features simultaneously.}
\label{vm33uprofil} 
\end{figure}

As we have seen in \citet{psa02} the morphological evolution of the orbits of
the family reaches a $y_{max}$ distance, beyond which the orbits grow
practically only in the $z$ direction. For building side-on profiles by using
the orbits of the x1-tree in 3D Ferrers bars this is an advantage, since it
restricts the extent of the peanut to a fraction of the bar length of the order
of half the distance to the end of the x1 bar, which is a desired property. For
our model this does not hold  in the case of the $mul$-periodic orbits that can
be used as alternative solutions for the orbital content of the peanuts. We use
again the orbits of the vm33u family to show this. In Fig.~\ref{vm33uprofil} we
plot together the side-on projections of seven periodic orbits of the vm33u
family at energies \ej $= -0.34$ to $-0.28$, all of which
are simple (vertical) unstable. In (a) we use orbits of one branch of the
family, while in (b) we consider both its branches, symmetric to the equatorial
plane. Ignoring the fact that these periodic orbits in our model are unstable,
so that only sticky-chaotic orbits could possibly be used to populate a peanut
structure, we observe that already at \ej$=-0.28$, the orbits contribute in
growing a profile faster in the $z-$ than in the $y-$direction. By including
more orbits at larger energies, or orbits from the branch of the characteristic
that goes backwards, towards the center of the system, we make the
$z_{max}/y_{max}$ ratio even larger as the projection of the orbits on the $y-$
axis shrinks. We remind that the longest x1, bar supporting orbit reaches a
4.1~kpc distance along the major axis, and that the \mbox{(\ej,$y$)} 
coordinates of the
$L_1$ Lagrangian point are $(-0.197,6.2)$. Thus, these orbits can support only 
features embedded
in a boxy bulge in the central parts of the model. At any rate, the profile of
the overlapping p.o. in (b) supports an overall boxy morphology, that harbours
simultaneously two kinds of X-features; one with branches passing through the
center of the system \citep[``CX'' in the terminology of][]{betal06} and another
one with wings extending in a direction vertical to the $y-$axis
\citep[characterized as ``OX'' by][]{betal06}. In galaxies, there are not known
cases in which the two types of X coexist. Also the almost vertical orientation
of the wings of the X is rather atypical. However, the stability and the extent
of the vm33u family in our model does not allow us to investigate the dynamical
mechanisms under which the appearance of just one of the X's could prevail in
profiles built by this family. 
 
\subsection{The importance of combining symmetric branches of orbits and
groups of orbits in a family for modelling morphological features:}

\begin{figure}
\begin{center}
\resizebox{80mm}{!}{\includegraphics[angle=0]{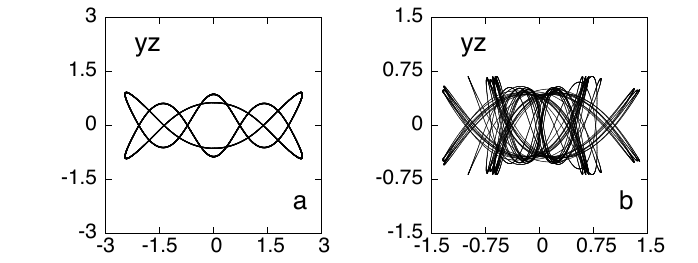}}
\end{center}
\caption{Structures that are formed in side-on projections, when we consider
pairs
of orbits, symmetric with respect to the equatorial plane. (a) The profile
formed by
the r\_tv1 family (Table~\ref{tab:mul2tab}). (b) A profile formed by a pair of
symmetric, quasi-periodic orbits trapped around a stable representative of vm35u
(Table~\ref{tab:mul3tab}).}
\label{symm} 
\end{figure}

At a given energy, as we move from the centre of the main island of a
simple-periodic orbit to its borders, we find in the surfaces of section groups
of smaller and smaller islands. Since our model is 3D, these groups of islands
are invariant tori belonging to the $mul$-periodic orbits described in our
tables
\citep[see figures 14 to 17 in][]{kpp11}. The existence of the $mul$-periodic
orbits creates in phase space a zone of morphological influence, because these
p.o. co-exist together with their symmetric, ``twin'', orbits  and their
unstable counterparts. Symmetry, often helps in building orbital structures.
Both ``frowns'' and ``smiles'' are needed to build the peanut by means of x1v1
and x1v1${^\prime}$ orbits. The plethora of orbits one can find in rotating
triaxial systems offers a lot of examples of such pairs. In Fig.~\ref{symm} we
give two of them.
In (a) we plot the side-on view of the r\_tv1 p.o. (see Table~\ref{tab:mul2tab})
together with its $z-$symmetric one, while in (b) we give a quasi-periodic
orbit, together with its symmetric with respect to the equatorial plane, trapped
around a stable representative of the vm35u family (Table~\ref{tab:mul3tab}) at
\ej$=-0.279$, which has  initial conditions $(-0.0229, -0.4088, 0.2086,
0.0865)$. The patterns formed in these profiles are promising for supporting an
X feature. Nevertheless, in order to decide about the effectiveness of a family
in building a structure, one needs to consider successive representatives of it
and see if the feature to be modelled appears in such composite profiles. The
most common example associated with the boxy bulges that describes the effect,
is the way the frown-smile simple-periodic x1v1 orbits support the X structure
in them. It is not the wings of the frown-smile x1v1 orbits that support it, but
the fact that their apocenters are aligned along rays, which are eventually the
branches of the X \citep[see figure 19a in][and especially figure 11 in
PKa]{psa02}. The rays are a structure that appears in the composite profiles and
not a morphological feature of the individual periodic orbits. Effects
appearing in the orbital profiles after combining orbits from symmetric branches
of a family, or after combining successive members of a family in an energy
range, i.e. building composite profiles as in \citet{psa02}, should be used as 
a criterion for qualifying or excluding this family from
being considered as a building block for a specific morphological feature.

\subsection{Side-on and face-on boxiness:}

None of the \textit{individual} face-on orbital patterns of the
$mul$-periodic orbits is observed in real galaxies or in density maps of
$N$-body simulations snapshots. For example, despite the fact that the rm21
orbits are found in all models in the orbital studies we cite, their morphology
does not appear in any galactic bar. However, when both $y-$symmetric orbits are
taken into account a box is formed (Fig.~\ref{boxy1}). This is in agreement with
the findings in PKb for the orbits in the vILR region that are responsible for
the inner boxiness and with the conclusions of \citet{cppsm17} for orbits
encountered at all \ej along their characteristics. Boxiness in general is
associated with quasi-periodic and sticky chaotic orbits close to the borders of
the x1 stability islands. This is exactly the region where we find
$mul$-periodic orbits. We have to note that the possibility of the appearance of
a morphological feature when sticky-chaotic orbits are integrated for long time
cannot be a priori excluded. In PKb, in \citet{tp15} and in \citet{cppsm17} such
orbits are proposed for explaining the \textit{face-on} X features that appears
in some galactic bars. In the present paper this happens e.g. with the orbits in
Fig.~\ref{rm213d}b,c (considering also their y-symmetric counterparts) if
integrated for long enough. However, since we did not find another similar case,
based on families with $n\geq 2$, in order to decide about the generality of
this dynamical mechanism, a model-dependent systematic study is needed.

For the edge-on profiles, We observe that the larger the multiplicity, the more
boxy the side-on view of the family is. The same holds for the \ej at which a
family is introduced in the system for all families of a certain multiplicity.
The larger the \ej at which it is bifurcated from x1 the $mul$-periodic family,
the more boxy is its side-on view. This is the reason why we stopped giving the
profiles of more orbits bifurcated from x1, as the stability curves of this
family vary (Fig.~\ref{x1stm2} and Fig.~\ref{x1stm3}). In general,
for p.o. of high multiplicity, and for those bifurcated at large \ej,
morphologies can be 
simply
characterized as complicated Lissajous patterns and an overall boxiness is the
only feature of their morphology.

\subsection{The complexity of the structure of phase space
increases as we approach corotation:}

The morphology of an orbit at a given energy depends on the families of resonant
periodic orbits that exist at that energy, on their stability that will
determine the phase space structure at their neighbourhood \citep[see][]{kp11,
kpc11, kpc13}, as well as on the location of the initial conditions of the orbit
in phase space. With increasing energy, i.e. moving from the center towards
corotation, we reach progressively more radial and vertical resonances, so new
families are introduced in the system and thus the structure of phase
space becomes more complex. Orbits with initial conditions that are close to
each other will probably follow totally different trajectories. This is an
expression of the presence of chaos close to corotation, as in 2D systems
\citep{gco81}. This hinders the growing of structures like bars at this region.

\subsection{The $mul$-periodic orbits are more important at the central parts
of the bars.}

The $mul$-periodic orbits we consider as building blocks of the orbital
patterns encountered in the papers we cite, are bifurcated at low energies. 
Thus they
are important for the inner morphology of the bars, at distances from the center
up to half the length of the semimajor axis. Most orbits given in our Tables
have sizes that in our model would contribute only in the central parts of the
bars. For example, orbits with simple shapes like those of the
vm31 family\footnote{The edge-on views of this family resemble a shape 
encountered already in orbits found in 
triaxial systems, called ``pretzel'' by \citet[][see their figure 4]{mv99}. We 
do not give to it any name here, to avoid confusion, since a similar term is 
used by PWG for the side-on shape we find in our vm33u orbits.}, if populated, 
would be found, in most models, embedded
in the central parts of the galaxies, not being able to affect the overall
morphology of a bar, neither face- nor edge-on. This is probably the reason that
this family is not traced in other papers, despite its promising shape,
especially if both $z-$symmetric branches are considered.

\subsection{$mul$-periodic bifurcations at high energies:}

We find that $mul$-periodic bifurcations of x1 and x1-tree families at high
energies, as well as the planar families beyond the 4:1 resonance gap
and its bifurcations at those energies, are in general not bar-supporting.
They correspond to regions of the bar beyond its inner thick part, or
beyond the end of the bar altogether. Their morphological evolution as \ej
increases is towards rounder shapes than the elliptical projections of the
x1-tree orbits on the equatorial plane. 

As a further, albeit less strong, obstacle for higher multiplicity p.o. to
determine
morphological profiles, is that higher multiplicity families have longer 
periods and thus would need longer times to impose their individual
morphologies in all three projections. Indeed, one can intuitively understand 
this by considering
that exactly at the bifurcating point the representatives of the mother and
child families are morphologically identical, while the period of the
``mother'' is T and the period of the ``child'' is $mul$T. As \ej increases
both T and $mul$T increase. As an example we give in Table~\ref{periods}
the periods of x1 and rm21 p.o. in the interval $-0.331<E_J<-0.312$, in which
rm21
exists in our model. Despite the fact that the T(rm21)/T(x1) ratio decreases
with the energy, it is always larger than one.
\begin{table}{}
\caption[]{The periods of x1 and rm21 families in the interval
$-0.331<E_J<-0.312$. We have always T(x1)$<$T(rm21).}
\label{periods}
\begin{tabular}{ccccccc}
\hline
\ej & $-0.331$ & $-0.33$ & $-0.325$ & $-0.32$ & $-0.315$ & $-0.312$ \\
T(x1)& 23.1 & 23.8 & 26.8 & 29.8 & 32.7 & 34.5 \\
T(rm21)& 46.2 & 46.8 & 49.3 & 51.5 & 53.5 & 54.3\\
\hline
\end{tabular}
\end{table}


\subsection{Model-dependency of the orbits and their role for supporting
specific
structures:}

The orbits we are presenting are not a special class of orbits existing in our
specific model or in rotating Ferrers bars. The orbits, each one with its own
characteristic morphology, are linked to the radial and vertical resonances of
any rotating, triaxial potential in autonomous Hamiltonian systems. Thus, the
patterns we observe in the Tables of this paper, as well as those of the
$1-$periodic orbits in \citet{spa02a}, have been encountered in several
different models that may include a Ferrers bar or not. Thus, they are
expected to exist also in models with a peanut-shaped bar. In this case their
relative importance for reproducing a peanut morphology should be estimated by
means of Schwarzschild-type, self-consistent models. Alternatively, one can use 
an N-body snapshot from which we can obtain the potential and forces, as well 
as all the individual orbits that constitute it. We will follow this latter 
alternative in future papers. 
 
\subsection{The role of chaotic orbits:}

We note that in many cases of orbits, the morphological patterns
encountered in orbital analyses correspond to families that are mainly 
\textit{un}stable. This can happen either because these families
are unstable over large $\Delta$\ej intervals, or even because they are
introduced as unstable and remain unstable as \ej varies. \citep[see e.g. the 
example 
described by][]{ph18}. If such families prevail in realistic models indeed, then
either the role of sticky chaotic orbits in building the bars (PKa,b) is much
more pronounced than appreciated until now, or we have mass distributions, where
the stability of the main families is different than in the known, analytic 
models.

\vspace{0.5cm}
The main point of the present paper was to indicate the origin of $mul$-periodic
orbits and discuss the obstacles they have to overcome in order to become
important for the overall morphology of galactic bars. We also compared their
morphology with orbital patterns found and presented in several relevant
works. In the subsequent papers of this series, we will investigate the relative
contribution of the orbits presented here in shaping the morphology of
specific 3D $N-$body bars.

\vspace{0.5cm} 
\noindent \textit{Acknowledgements}

We thank Prof. G.~Contopoulos for fruitful discussions and very useful comments.
This work has been partially supported by the Research Committee of the Academy
of Athens through the project 200/895. 
PAP thanks the financial support of Aix Marseille Universit\'{e} for a 2-month
visiting professorship at LAM.

\appendix\section{Quasi- and non-periodic orbits}
\label{sec:qporbs}
 
In Sections~\ref{sec:2po} and \ref{sec:3po} we have indicated the origin of the 
main 2-, 3-, 4-, 5- and 6-periodic bifurcations of the planar x1 family, while 
in Section~\ref{sec:highej} we examined the role of $mul$-periodic orbits with 
large energies in determining the structure of the bar. The structure of these 
orbits has to be compared with the morphology of orbits encountered in the 
relevant literature, which are clearly not periodic (or at least not 
$mul$-periodic with a small $mul$). The peanut-supporting orbits depicted in 
the 
papers we mentioned in the introduction, point either to quasi-periodic, or to 
sticky-chaotic orbits behaving like regular for a reasonably long time.

Periodic orbits are mathematical objects and can be considered as the backbone 
of the phase space in a system. However, in order to follow orbits that are
more 
frequently found in realistic barred galaxy potentials, those of real galaxies, 
or of simulated ones, one has to integrate initial conditions away from those 
that exactly correspond to a periodic one. The periodic orbits structure the 
phase space and their presence influences the morphology of other quasi- or 
non-periodic orbits at the same energy. In 2D models for galactic bars we have 
described this already in \citet{paq97} for orbits that support the outer 
boxiness of the bars. The orbital patterns we had found in that paper were 
hybrid morphologies of existing periodic orbits, which can be stable or 
unstable. 

One can get a fair impression of the possible morphologies that can be supported
at a given energy by simply integrating directly a number of initial conditions
on a dense grid imposed on a 2D surface of section. This gave useful results in
a number of studies \citep[see e.g.][]{pkg10, tp15}. Unfortunately, and contrary
to 2D models, in 3D systems, starting perturbing planar orbits for example in
$(x,p_x)$ surfaces of section, can only reveal a subset of the morphologies that
can be encountered. The Poincar\'{e} surfaces of section are 4D and so, the
initial conditions of a single p.o. are subject to perturbations in four
directions. Constructing a dense 4D grid will lead to a huge amount of data and
this is not easily manageable. 

The information we get by inspecting a 2D surface of section in a 2D model is 
not available in the 4D surfaces of section in 3D autonomous
Hamiltonian systems. It 
is not only that we cannot directly classify as regular or chaotic an orbit by 
simple inspection of the location of its initial condition on the surface of 
section, but it is mainly the inability to trace the relative location of the 
stability islands around a stable p.o. in the chaotic seas, in which they are 
embedded. Thus, it is difficult to delimit zones of morphological influence of a 
p.o. in phase space. Although we could visualize 4D invariant tori of stable 
p.o. \citep{pz94, kp11} and in some cases the manifolds associated with unstable 
p.o. \citep{kpc11, kpc13} in the 4D spaces of section, we could only in few 
specific cases navigate ourselves in regions of the 4D space, where several 
islands of periodic orbits co-exist \citep[][PKa, KPP]{kpc13}. Nevertheless, in 
the present study we disturbed a large number of characteristic planar
and 3D 
$mul-$periodic orbits and we succeeded in finding quasi- and non-periodic
orbits 
with morphologies pointing to structures that could be related to the peanut or 
X-shaped structures (e.g. frown-smile shapes), which we encounter in 3D bars.

Even though perturbing p.o. in a non-systematic way does not provide all the
information that a systematic search would, it can still be most useful. The 
first thing that one can estimate is the easiness or difficulty with which some
patterns appear in the three projections of orbits in different models. As
expected, the patterns that one encounters easier are associated with the
presence of 1-periodic orbits, since the latter are found in the centres of
stability islands that occupy large volumes in the phase space. In barred galaxy
models these orbits are the x1 family and its 1-periodic 3D
bifurcations that build the
x1-tree \citep{spa02a}, which have elliptical shapes in the $(x,y)$ plane.
Three-dimensional, non-periodic orbits with elliptical projections on the
equatorial plane can be found either by perturbing quasi-periodic orbits
corresponding to invariant curves \textit{close to} x1 in $(x,p_x)$ surface of
section, or by perturbing one or more of the four initial conditions of the 3D
orbits belonging to the families of the x1 tree (x1v1, x1v2 etc.). 
Independently of their origin and their regular or sticky-chaotic character, the
supported elliptical morphologies on the equatorial plane are almost identical
and are ubiquitous in papers about orbits in barred galaxy models (e.g. the
orbits in figure 6, 3rd row in AVSD; the orbits in figure 3, 1st row and the
orbits in figure 10 in \citet{cppsm17}; the orbits in figures 6, 1st row and
figure 7, 1st and 2nd rows, in GLA; the orbits in figures 3 and 4 in KPP; the
orbit in figure 15 in PKa; the orbits in figure 1, three first rows, in figure
3, 1st row, right, and figure 4, 4th and 5th row in VSAD; the orbits in figures
10 and 11 in WM-D etc.). In order to avoid elliptical shapes in the $(x,y)$
projections of the orbits, one has to deviate more from the initial conditions
of the x1 or x1-tree orbits and consider orbits close to the edges of the
stability islands (both at their ``regular'' and their ``chaotic'' side). This
can lead to single (only in one projection), or double (both in their face-on
and edge-on views) boxiness of orbits of galactic bars \citep{cppsm17} and can
be particularly important for the structure of the peanuts when we consider
orbits in the vILR resonance region (PKb). 

\begin{figure}
\begin{center}
\resizebox{80mm}{!}{\includegraphics[angle=0]{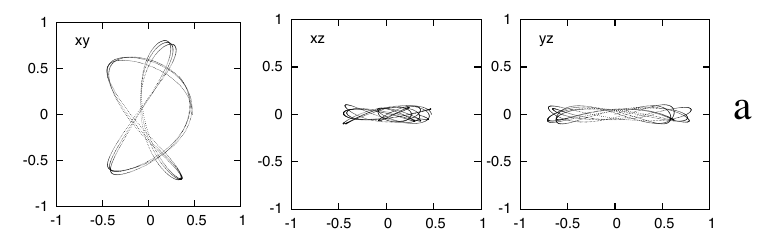}}
\resizebox{80mm}{!}{\includegraphics[angle=0]{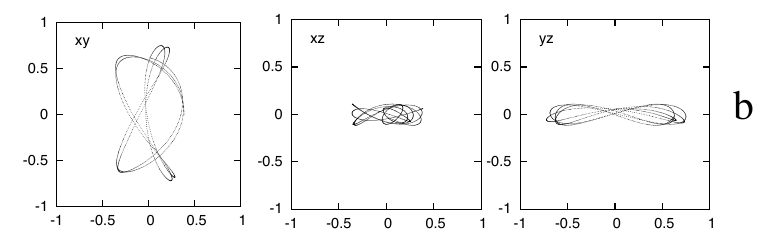}}
\resizebox{80mm}{!}{\includegraphics[angle=0]{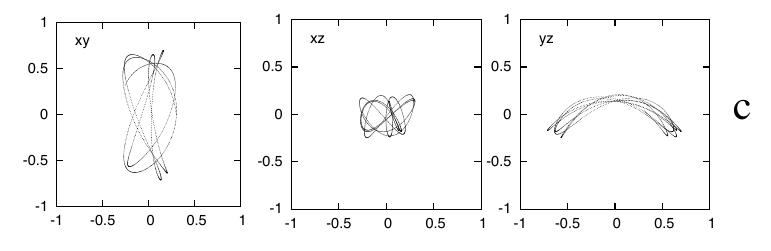}}
\end{center}
\caption{
Three 3D orbits that have in their face-on projection an rm21-like            
morphology. This is combined in (a) with a boxy side-on view, in (b) with a
x1v2-like,
and in (c) with a x1v1-like one.}
\label{rm213d} 
\end{figure}

Depending on the various resonant families that exist at a given energy, the 
face-on view of a quasi- or non-periodic orbit can be combined with different 
side-on projections. We give as an example three 3D orbits with face-on views 
pointing directly to the planar rm21 family (Fig.~\ref{rm213d}). As we already 
commented, the frequency with which the rm21 morphology appears in orbits 
presented in the orbital studies of rotating barred potentials is large (see 
section \ref{params} above). This is due to the fact that it exists in 
considerable energy ranges in the models, close to the vILR region, and its 
stability islands, together with those of rm22, always surround the very
central invariant curves around x1, forming 
a zone of morphological influence in this area (see e.g. Fig.~\ref{ofmainm2}).
By 
inspecting of surfaces of section like the one in Fig.~\ref{ofmainm2}, we 
realize e.g. that orbits with a $|\Delta p_x \gtrapprox 0.1|$ from the periodic 
orbit, enter the zone of influence of the rm21 and rm22 families. This can 
easily happen if the dispersion of velocity is locally slightly increased in a 
model. In other cases of stronger bars, there is also a sticky zone with orbits 
of similar to rm21, rm22 and rm21u, rm22u morphologies around the x1 stability 
island (a characteristic case is given in figure 1 in PKb). 

By perturbing these regular and/or sticky chaotic orbits in the vertical
direction we reach orbits that retain until a certain value of the perturbation
the rm21 face-on morphology. However, they differ between them in their edge-on
projections. The latter depend on the location of the $z, p_z$ initial
conditions in the 4D space of section, which we use. It is difficult to predict
a priori how the morphology of quasi-periodic orbits, or of orbits on manifolds,
changes as we move along a certain direction in phase space. Until now it has
been described only in particular cases, such as the case of the x1, x1v1 and
x1v2 orbits in \citet{kpc13} and in PKb. In general we could say that the
morphology is influenced by the proximity of the initial conditions of an orbit
to those of other periodic orbits at the same energy. The orbits in
Fig.~\ref{rm213d} are integrated for a few rm21 periods and have in (a)
\ej$=-0.32396$ with initial conditions $(0.475, 0, 0, 0.05)$, in (b)
\ej$=-0.3268$ and $(0.385, 0.06, 0, 0)$ and in (c) \ej$=-0.3268$ and $(0.3,
0.14, -0.02, 0.03)$. 
We can clearly see the similarity of their $(x,y)$ projections with the orbits 
of the rm21 family, while the side-on views are different. The side-on view of 
the orbit in Fig.~\ref{rm213d}a can be characterized just ``boxy'' already just 
after three rm21 orbital periods, while those of Fig.~\ref{rm213d}b and 
Fig.~\ref{rm213d}c are similar for example with those in figure 7, 4th row and 
figure 7, 3rd row in GLA, respectively. They have a x1v2-like 
(Fig.~\ref{rm213d}b) and a x1v1-like (Fig.~\ref{rm213d}c) side-on morphology. 
Also in WAM (figure 5, 2nd and 3rd column) we encounter orbits with similar 
combinations of face- and side-on profiles as in Fig.~\ref{rm213d}b and c. We 
can find more edge-on profiles combined with rm21 face-on morphologies, if we 
consider also quasi- and non-periodic orbits associated with its 3D 
bifurcations, as e.g. the rm21\_vm4 orbit given in Table~\ref{tab:mul2tab} or 
the orbit in figure 2 in \citet{kpp11}.

\begin{figure}
\begin{center}
\resizebox{80mm}{!}{\includegraphics[angle=0]{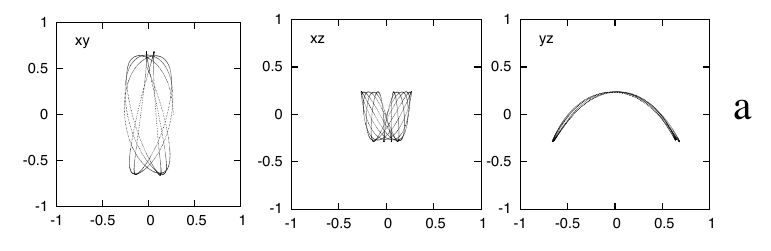}}
\resizebox{80mm}{!}{\includegraphics[angle=0]{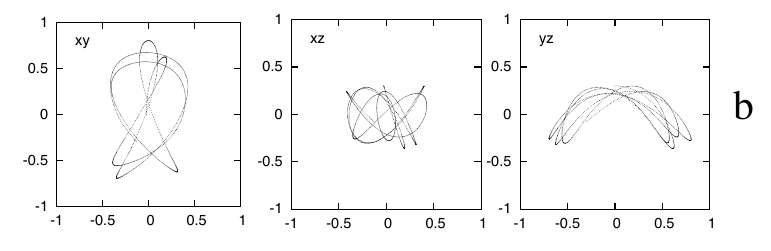}}
\resizebox{80mm}{!}{\includegraphics[angle=0]{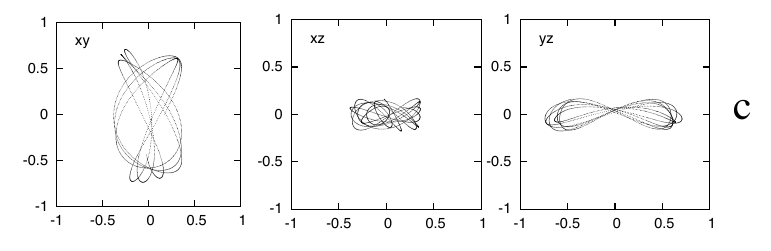}}
\end{center}
\caption{Face-on rm21u and rm22u morphologies combined with side-on x1v1-like
(a) and (b),
and x1v2-like shapes (c).}
\label{rm21u3d} 
\end{figure}

In reverse, basing ourselves on the edge-on views, at energies larger than those 
of the vILR region, the 3D orbits that have side-on profiles pointing to the 
x1v1 and x1v2 families are ubiquitous. Besides the quasi-periodic orbits of 
these two families (with elliptical or boxy face-on projections) and their 
associated sticky orbits, frown - smiles and ``$\infty$''-like edge-on profiles 
can be found combined with various face-on morphologies. The morphology of the 
planar unstable families rm21u and rm222u, bifurcated together with rm21 and 
rm22, can also support the known profiles of the vertical families bifurcated at 
the vILR. Examples are given in Fig.~\ref{rm21u3d}. In (a) we have an orbit with 
an rm21u-like face-on morphology with \ej=$-0.32281$, and initial conditions 
$(0.266, 0.24, 0, 0)$, in (b) a rm21u-like orbit with $-0.318, (-0.026, 0.3, 0, 
0)$ and in (c) a rm22u-like orbit with $-0.3268, (0.168, 0.07, 0.12, 0)$ 
respectively (the rm21u and rm22u families have the same morphology, and 
symmetric with respect to the x-axis). The combination of face-on and side-on 
views of the orbit in Fig.~\ref{rm21u3d}b is encountered also in figure 7, 5th 
row in GLA, while different side-on views in combination with an rm21u-like 
face-on morphology are found in orbits mentioned at the last column of the rm21u 
entry in Table~\ref{tab:mul2tab}. The rm21u and rm22u are the unstable p.o. we 
find between the rm21 and rm22 stability islands (Fig.~\ref{ofmainm2}). Thus, 
their morphology is also a common one.
 
Besides the x1v1 and x1v2 edge-on profiles, the next common morphology that is
encountered in the various models, is a boxy one, with sharply defined edges,
but without a particular morphology inside the box. The mechanism that favours
the appearance of such a structure, instead of a particular edge-on shape, is
presented in PKb and in \citet{cppsm17}. Apart from the cases that are presented
in these two papers, many examples of boxy orbits can be found in the relevant
literature, as e.g. in AVSD (figure 6, 1st and 4th row), \citet{dvm11} (figure
11), GLA (figure 6, two last rows), VSAD (all orbits in figure 1 and the orbits
in figure 4, three first rows), WAM (figure 21, first and last columns).

\begin{figure}
\begin{center}
\resizebox{80mm}{!}{\includegraphics[angle=0]{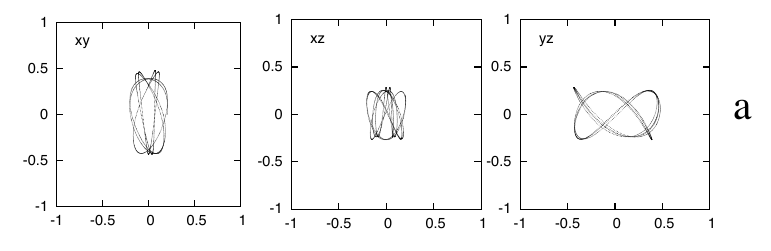}}
\resizebox{80mm}{!}{\includegraphics[angle=0]{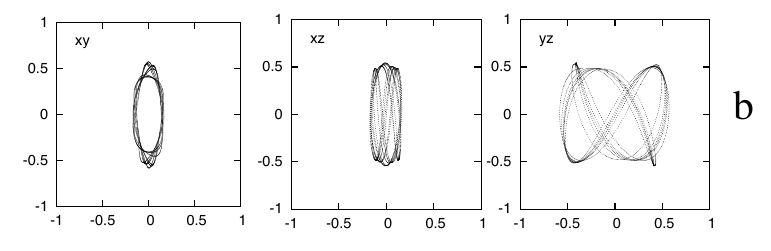}}
\resizebox{80mm}{!}{\includegraphics[angle=0]{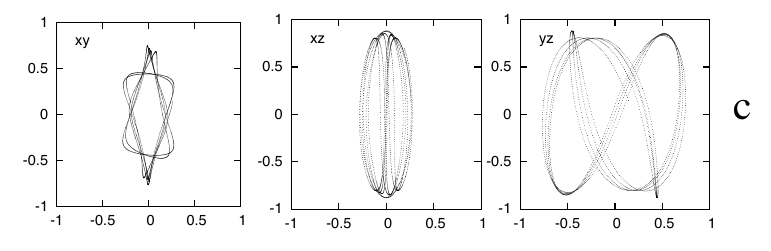}}
\end{center}
\caption{Three profiles of 3D orbits with vm33u-like side-on views, but with
considerably different face-on views. The energies of the orbits increase from
$-0.333$ (a), to $-0.31$ (b), and to $= -0.28$ (c). We observe that the
$z_{max}/y_{max}$ ratio increase with the energy.}
\label{oqvm33} 
\end{figure}

\begin{figure}
\begin{center}
\resizebox{80mm}{!}{\includegraphics[angle=0]{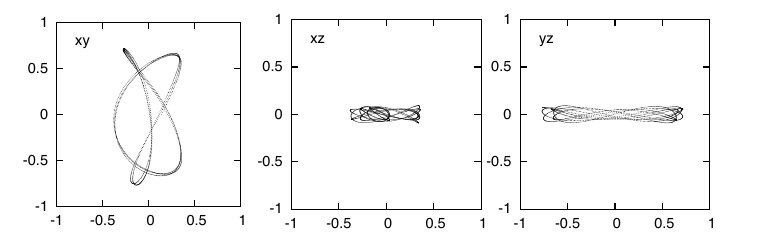}}
\end{center}
\caption{An orbit with initial conditions on an rm21 stability island perturbed 
in
the vertical direction, at \ej$=-0.3268$.}
\label{rm21z} 
\end{figure}

Coming to discuss the side-on profiles in Fig.~\ref{oqvm33}, we need to 
specifically mention the side-on profiles given in figure 2 in PWG. While the 
orbits in panels E and F of their figure 2 point to sticky chaotic and 
quasi-periodic, x1v1-like morphologies respectively (with any kind of face-on 
projection they may be combined), those in panels A to D are almost absent from 
the rest of the orbital studies we cite. An exception is the profile in panel C, 
emphasized in their figure 5 bottom, which appears also in figure 6, last row in 
AVSD, in figure 4, last row in VSAD and in figure 21 (2nd and 3rd columns) in 
WAM. The side-on morphology of all these orbits points directly to vm33u, the 
unstable, 3-periodic vertical bifurcation of x1 at its second tangency  with the 
$b=-2$ axis, when it is also considered to be 3-periodic (see 
Table~\ref{tab:mul3tab}). Three non-periodic orbits that retain the side-on 
vm33u morphology are given in Fig.~\ref{oqvm33}. 
The orbits are vertically unstable (the vertical index $b2=-5.3$ at \ej $\approx 
-0.29$), so the depicted side-on structure is preserved only when small radial 
deviations from the initial conditions of the p.o. are applied. In 
Fig.~\ref{oqvm33}a we have \ej $= -0.333$ and $(0.2, 0.203, 0.02, -0.082)$. In 
Fig.~\ref{oqvm33}b we have respectively $-0.31$ and $(0.15, 0.398, 0.0182, 
-0.137)$, which are the initial conditions of the periodic orbit at the same 
energy, differing only by $\Delta x_0 = 0.0074$. The orbit is integrated for 
about three orbital periods of the periodic orbit. By continuing the integration 
for two more periods, the morphology becomes boxy. In Fig.~\ref{oqvm33}c, 
\ej\!$= -0.28$ and the initial conditions $(0.19 0.695, 0.0486, -0.1396)$. In 
this case we observe, that the $z$-dimension of the orbit is larger than the one 
along $y$. The fact that the side-on views of the orbits reach a maximum length 
(i.e. a maximum extent along the $y$ axis) at a given energy, beyond which they 
practically increase only in height, is a known property of the 3D orbits in 
Ferrers bars. It has been discussed in \citet{psa02} in relation to the extent 
of the x1v1 peanut and we will also return to this in Section~\ref{sec:concl}. 
The family becomes stable in our model at even larger energies, where the 
$z_{max}/y_{max}$ ratio of the orbits is even larger. The face-on view of the 
orbit combined with the vm33u profile given by PWG in their figure 5, top, 
belongs to the rm21-like patterns, with an asymmetry in its loops that is 
typical for many of the orbits we encountered. We give in Fig.~\ref{rm21z} an 
example of an orbit with \ej$=-0.3268, (0.138, 0.05, 0.12, 0)$, for which we 
take initial conditions on a stability island in the $(x,px)$ plane of section 
belonging to rm21 and we add $z=0.05$ in its initial conditions. So, here we 
have one more example of different morphological combinations in different 
projections.

\begin{figure}
\begin{center}
\resizebox{80mm}{!}{\includegraphics[angle=0]{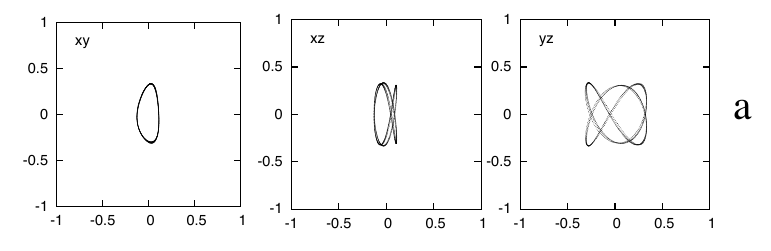}}
\resizebox{80mm}{!}{\includegraphics[angle=0]{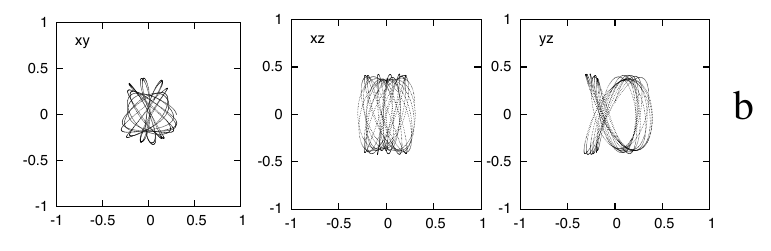}}
\resizebox{80mm}{!}{\includegraphics[angle=0]{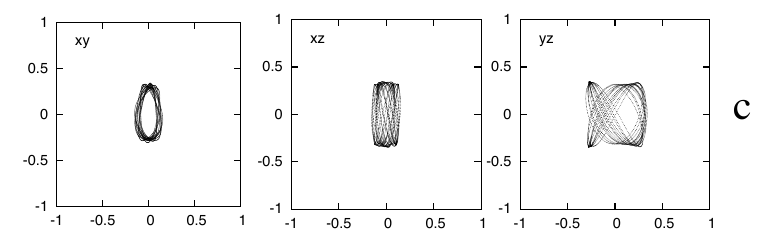}}
\end{center}
\caption{Three examples of 3D orbits reproducing the side-on profile of the 
unstable
family x1mul2u (vm21u). In (a) and (c) they are for energies $-0.333$ and
$-0.33307$, while in (b) for $-0.325$.}
\label{x1mul2uqz} 
\end{figure}

We find similar results for the profile in panel A in figure 2 in PWG. It is the 
family bifurcated from x1, at the same energy as x1mul2, but as unstable. 
Despite the fact that we did not encounter a similar side-on profile in other 
studies, we note that it is easily obtained by perturbing orbits of the stable 
family (x1mul2), underlying the fact that the initial conditions of the 
representatives of the two families at a given energy are close. In 
Fig.~\ref{x1mul2uqz} we give three such characteristic orbits. In (a) we have an 
orbit with \ej=$-0.333$ and initial conditions $(0.108, 0.294, -0.0066, 
0.0537)$, in (b) $-0.325$ and $(0.3, 0.13, -0.007, 0.2)$, while in (c)  
$-0.33307$ and $(0.08, 0.3, 0, -0.05)$ respectively. Here we have a situation 
similar to the one presented by \citet{ph18} for the x1v1/x1v2 pair, where at a 
given energy side-on profiles can be easily built either by quasi-periodic or by 
sticky chaotic orbits.

\begin{figure}
\begin{center}
\resizebox{80mm}{!}{\includegraphics[angle=0]{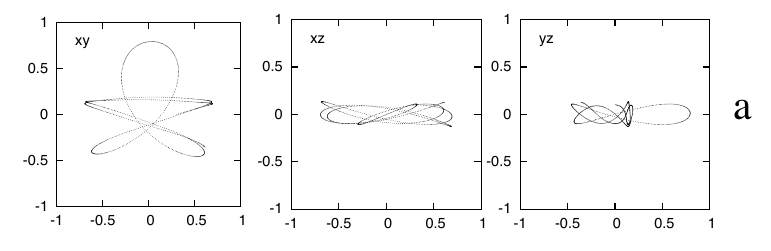}}
\resizebox{80mm}{!}{\includegraphics[angle=0]{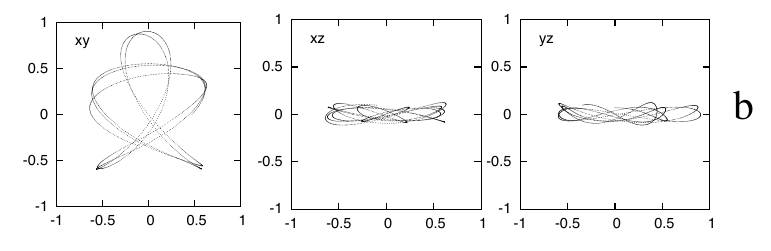}}
\end{center}
\caption{Two examples of 3D orbits with similar lengths along the $x$ and $y$ 
axes.
They are typical in bar orbital studies and may contribute in 
building the bar
central parts.}
\label{square} 
\end{figure}

Finally, we note that there is a class of non-periodic orbits, well represented 
in papers with orbital studies, with a characteristic shape in their 
face-on views. Most of them have morphologies similar to rm21u at energies much 
larger than the one at which this family introduced in the system. Two examples 
are 
given in Fig.~\ref{square}.
The orbit in (a) has \ej$=-0.319$ and initial conditions $(0.306, 0.1, 0.23,
0)$, while the one in (b) $-0.318$ and $(0.47, 0.07, 0.13, 0)$ respectively.
Their face-on structure is similar to the one of orbits like the one in figure
11, middle row in \citet{dvm11}, the orbit in figure 3, 3rd row right, in VSAD
or the orbit in figure 5, left column, in WAM. These orbits reflect the
morphological evolution of the rm21 and rm22 pair of families as energy
increases. Such orbits, not being elongated along the major axis of the bar can
populate only the central parts of galactic bars, and can be associated
with the
thick part of the bar.

\label{lastpage}


\begin{thebibliography}{}
\bibitem[\protect\citeauthoryear{Abbot et al.}{2017}]{avsd17} Abbott C. G.,
Valluri M., Shen J., Debattista V. P., MNRAS 470,
1526 (AVSD)
\bibitem[\protect\citeauthoryear{Athanassoula}{1992}]{a92} Athanassoula E.,
1992, MNRAS 259, 328
\bibitem[\protect\citeauthoryear{Athanassoula}{2016}]{a16} Athanassoula E, 2016,
in Astrophysics and Space Science Library
418, Galactic Bulges, E. Laurikainen, R. Peletier, D. Gadotti (eds.),
Springer International Publishing, Switzerland, p. 391
\bibitem[\protect\citeauthoryear{Athanassoula \& Beaton}{2006}]{ab06}
Athanassoula E., Beaton R. 2006, MNRAS, 370, 1499
\bibitem[\protect\citeauthoryear{Athanassoula et al.}{1983}]{a83} Athanassoula
E.,
Bienayme O., Martinet L., Pfenniger D. 1983, A\&A 127, 349
\bibitem[\protect\citeauthoryear{Athanassoula et al.}{2015}]{a15} Athanassoula
E., Laurikainen E., Salo H., Bosma A.  2015, MNRAS, 454, 3843
\bibitem[\protect\citeauthoryear{Athanassoula et al.}{2010}]{a10} Athanassoula
E., Romero-G\'omez M., Bosma A., Masdemont J. 2010, MNRAS, 407, 1433
\bibitem[\protect\citeauthoryear{Broucke}{1969}]{b69} Broucke R., 1969, NASA
Techn. Rep. 32, 1360
609
\bibitem[\protect\citeauthoryear{Bureau et al.}{2006}]{betal06} Bureau M.,
Aronica
G., Athanassoula E., Dettmar R.-J., Bosma A., Freeman K. C., 2006, MNRAS, 370,
753
\bibitem[\protect\citeauthoryear{Chaves-Velasquez et al.}{2017}]{cppsm17}	
Chaves-Velasquez L., Patsis P. A., Puerari I., Skokos C., Manos T. 2017, ApJ
850, 145
\bibitem[\protect\citeauthoryear{Chaves-Velasquez et al.}{2019}]{cppmp19}	
Chaves-Velasquez L., Patsis P. A., Puerari I., Moreno E., Pichardo B., 2019,
ApJ in press (arXiv:1812.04068)
\bibitem[\protect\citeauthoryear{Contopoulos}{1981}]{gco81} Contopoulos G.,
1981, A\&A 102, 265
\bibitem[\protect\citeauthoryear{Contopoulos}{1983}]{gco83} Contopoulos G.,
1983, ApJ 275, 511
\bibitem[\protect\citeauthoryear{Contopoulos}{1986}]{gco86} Contopoulos G.,
1986, Celest. Mech. 38, 1
\bibitem[\protect\citeauthoryear{Contopoulos}{1988}]{c88} Contopoulos G.,
1988, A\&A 201, 44 
\bibitem[\protect\citeauthoryear{Contopoulos}{2004}]{gcobook} Contopoulos G.,
 2004, "Order and Chaos in Dynamical Astronomy", Springer-Verlag, Berlin,
Heidelberg,
New York
\bibitem[\protect\citeauthoryear{Contopoulos \& Grosb{\o}l}{1989}]{cg89}
Contopoulos G. \& Grosb{\o}l P., 1989, A\&ARv, 1,
261
\bibitem[\protect\citeauthoryear{Contopoulos \& Harsoula}{2008}]{ch08}
Contopoulos G. \& Harsoula M., 2008, Int. J. Bifurc. Ch. 18, 2929
\bibitem[\protect\citeauthoryear{Contopoulos \& Magnenat}{1985}]{cm85}
Contopoulos G., Magnenat P., 1985, Celest. Mech. 37, 387
\bibitem[\protect\citeauthoryear{Contopoulos \& Papayannopoulos}{1980}]{cp80}
Contopoulos G. \& Papayannopoulos T., 1980, A\&A 92, 33
\bibitem[\protect\citeauthoryear{Deibel et al.}{2011}]{dvm11} Deibel A.T.,
Valluri
M., Merritt D., 2011, ApJ 728, 128
Elmegreen
\bibitem[\protect\citeauthoryear{Erwin \& Debattista}{2013}]{ed13} Erwin P.,
Debattista V. P., 2013, MNRAS, 431, 3060

\bibitem[\protect\citeauthoryear{Ferrers}{1877}]{f887}
Ferrers, N. M. 1877, Quart.J.Pur.Appl.Math., 14, 1
\bibitem[\protect\citeauthoryear{Gajda et al.}{2016}]{ggla16} Gajda G.,
\L{}okas E. L., Athanassoula E. 2016, ApJ 830,
108 (GLA) 
\bibitem[\protect\citeauthoryear{Hadjidemetriou}{1975}]{hdj75} Hadjidemetriou
J., 1975, Celest. Mech., 12, 255
\bibitem[Katsanikas \& Patsis (2011)]{kp11} Katsanikas M., Patsis P.A., 2011, 
  Int. J. Bif. Ch. 21-02, 467
\bibitem[Katsanikas et al. (2011b)]{kpp11} Katsanikas M., Patsis P.A.,
Pinotsis A.D., 2011b, Int. J. Bif. Ch. 21-08, 2331 (KPP)
\bibitem[Katsanikas et al. (2011)]{kpc11} Katsanikas M., Patsis P.A.,
Contopoulos G., 2011, Int. J. Bif. Ch. 21, 2321
\bibitem[Katsanikas et al. (2013)]{kpc13} Katsanikas M., Patsis P.A.,
Contopoulos G., 2013, Int. J. Bif. Ch. 23-02, 1330005
\bibitem[\protect\citeauthoryear{Machado \& Manos}{2016}]{mm16}	Machado R.E.G.,
Manos T., 2016, MNRAS 458, 3578
\bibitem[\protect\citeauthoryear{Manos \& Athanassoula}{2011}]{ma11} Manos T.,
Athanassoula E., 2011, MNRAS 415, 629
M.,
\bibitem[\protect\citeauthoryear{Merritt \& Valluri}{1999}] {mv99} Merritt D., 
Valluri M., 1999, AJ 118, 1177
\bibitem[\protect\citeauthoryear{Miyamoto \& Nagai}{1975}]{mn75} Miyamoto M.,
Nagai R., 1975, PASJ 27, 533
Springer
\bibitem[\protect\citeauthoryear{Patsis \& Zachilas}{1990}]{pz90} Patsis P. A.,
Zachilas L., 1990 A\&A 227, 37
\bibitem[\protect\citeauthoryear{Patsis \& Grosb{\o}l}{1996}]{pg96} Patsis P.
A.,
Grosb{\o}l P., 1996, A\&A, 315, 371
\bibitem[\protect\citeauthoryear{Patsis et al.}{1997}]{paq97} Patsis P. A.,
Athanassoula E., Quillen A. C., 1997, ApJ, 483, 731
\bibitem[\protect\citeauthoryear{Patsis et al.}{2002a}]{pags02} Patsis P. A.,
Athanassoula E., Grosb{\o}l P., Skokos Ch., 2002, MNRAS 335,
1049
\bibitem[\protect\citeauthoryear{Patsis et al.}{2002b}]{psa02} Patsis P. A.,
Skokos Ch., Athanassoula E., 2002, MNRAS 337, 578
\bibitem[\protect\citeauthoryear{Patsis \& Harsoula}{2018}]{ph18}  Patsis P.
A., Harsoula M., 2018, A\&A  612, 114
\bibitem[\protect\citeauthoryear{Patsis et al.}{2010}]{pkg10} Patsis P. A.,
Kalapotharakos C., Grosb{\o}l P., 2010, MNRAS 408, 22
\bibitem[\protect\citeauthoryear{Patsis \& Katsanikas}{2014a}]{pk14a}  Patsis P.
A., Katsanikas M., 2014, MNRAS 445, 3525 (PKa)
\bibitem[\protect\citeauthoryear{Patsis \& Katsanikas}{2014b}]{pk14b}  Patsis P.
A., Katsanikas M., 2014, MNRAS 445, 3546 (PKb)
\bibitem[Patsis \& Zachilas (1994)]{pz94}  Patsis P.A., Zachilas L., 1994, Int.
J. Bif. Ch. 4, 1399 
\bibitem[\protect\citeauthoryear{Pfenniger}{1984}]{pf84} Pfenniger D., 1984,
A\&A 134, 373
\bibitem[\protect\citeauthoryear{Plummer}{1911}]{pl11} Plummer H.C., 1911,
MNRAS 71, 460
\bibitem[\protect\citeauthoryear{Poincar\'{e}}{1899}]{poin99} Poincar\'{e} H.,
1899, Les Methodes Nouvelles de la Mechanique Celeste, Vol. III,
Gauthier-Villars,
Paris
\bibitem[\protect\citeauthoryear{Portail et al.}{2015}]{pwg15} Portail,
M., Wegg C., Gerhard O., 2015, MNRAS 450L, 66
(PWG)
\bibitem[\protect\citeauthoryear{Heisler et al.}{1982}]{hms} Heisler J.,
Merritt D., Schwarzschild M., 1982, ApJ 258, 490
\bibitem[\protect\citeauthoryear{Skokos et al.}{2002a}]{spa02a} Skokos Ch.,
Patsis P. A., Athanassoula E., 2002a, MNRAS 333, 847
\bibitem[\protect\citeauthoryear{Skokos et al.}{2002b}]{spa02b} Skokos Ch.,
 Patsis P. A., Athanassoula E., 2002b, MNRAS 333, 861
\bibitem[\protect\citeauthoryear{Tsigaridi \& Patsis}{2015}]{tp15} Tsigaridi
 L., Patsis P.A., 2015, MNRAS,  448, 3081
\bibitem[\protect\citeauthoryear{Valluri et al.}{2016}]{vsad16} Valluri M.,
Shen J., Abbott C. G., Debattista V. P., ApJ 818, 141 (VSAD)
\bibitem[\protect\citeauthoryear{Wang et al.}{2016}]{wam16} Wang Y.,
Athanassoula E., Mao S., 2016, MNRAS 363, 3499 (WAM) 
\bibitem[\protect\citeauthoryear{Wozniak \& Michel-Dansac}{2009}]{wd09} Wozniak
H., Michel-Dansac L., 2009. A\&A 494, 11 (WM-D)
\end{thebibliography}
\end{document}